\documentclass[runningheads]{llncs}

\usepackage{eccv}

\usepackage{eccvabbrv}

\usepackage{graphicx}
\usepackage{booktabs}

\usepackage[accsupp]{axessibility}  

\usepackage{hyperref}

\usepackage{orcidlink}

\usepackage{multirow}
\usepackage{multicol}
\usepackage{tikz}
\usepackage{amsfonts}
\usepackage{nicefrac}
\usepackage{microtype}
\usepackage{lipsum}
\usepackage{graphicx}
\usepackage{float}
\usepackage{algorithm}
\usepackage{algpseudocode}
\usepackage{booktabs}
\usepackage{makecell}
\usepackage{colortbl}
\usepackage{mathabx}
\usepackage{contour}
\usepackage{wrapfig}
\usepackage[normalem]{ulem}
\usepackage{soul}
\usepackage{bbm}

\usepackage{amsmath,amsfonts,bm}
\usepackage[english]{babel}

\def\eqref#1{equation~\ref{#1}}

\def\1{\bm{1}}

\DeclareMathAlphabet{\mathsfit}{\encodingdefault}{\sfdefault}{m}{sl}
\SetMathAlphabet{\mathsfit}{bold}{\encodingdefault}{\sfdefault}{bx}{n}

\newcommand{\R}{\mathbb{R}}

\newcommand{\commentsymbol}{$\triangleright$}{}
\algrenewcommand\algorithmiccomment[1]{\hfill \commentsymbol{} #1}
\makeatletter
\newcommand{\LineComment}[2][\algorithmicindent]{\Statex \hspace{#1}\textcolor{blue}{\commentsymbol{} #2}}
\makeatother

\definecolor{Gray}{gray}{0.91}
\newcolumntype{g}{>{\columncolor{Gray}}c}
\newcolumntype{G}{>{\columncolor{Gray}}r}

\newcommand{\Tech}{{{\sc UNIT}{}}}

\begin{document}

\title{\Tech{}: Backdoor Mitigation via Automated Neural Distribution Tightening}
\titlerunning{\Tech{}: Automated Neural Distribution Tightening}

\author{
Siyuan Cheng\inst{1}$^*$ \and Guangyu Shen\inst{1}$^*$ \and Kaiyuan Zhang\inst{1} \and Guanhong Tao\inst{1} \and \\
Shengwei An\inst{1} \and Hanxi Guo\inst{1} \and Shiqing Ma\inst{2} \and Xiangyu Zhang\inst{1}
}

\authorrunning{S.~Cheng, G.~Shen et al.}

\institute{Purdue University, West Lafayette, IN, 47906, USA
\email{\{cheng535,shen447,zhan4057,taog,an93,guo778,xyzhang\}@cs.purdue.edu}
\and
University of Massachusetts at Amherst, MA, 01003, USA \\
\email{shiqingma@umass.edu} \\
\textit{$^*$ denotes equal contribution}
}

\maketitle

\begin{abstract}
Deep neural networks (DNNs) have demonstrated effectiveness in various fields.
However, DNNs are vulnerable to backdoor attacks, which inject a unique pattern, called trigger, into the input to cause misclassification to an attack-chosen target label.
While existing works have proposed various methods to mitigate backdoor effects in poisoned models, they tend to be less effective against recent advanced attacks. In this paper, we introduce a novel post-training defense technique \Tech{} that can effectively eliminate backdoor effects for a variety of attacks.
In specific, \Tech{} approximates a unique and tight activation distribution for each neuron in the model. It then proactively dispels substantially large activation values that exceed the approximated boundaries.
Our experimental results demonstrate that \Tech{} outperforms 7 popular defense methods against 14 existing backdoor attacks, including 2 advanced attacks, using only 5\% of clean training data. \Tech{} is also cost efficient.
The code is accessible at \href{https://github.com/Megum1/UNIT}{https://github.com/Megum1/UNIT}.
\end{abstract}

\section{Introduction} \label{sec:intro}
As deep learning (DL) continues to reshape industries, spanning from transportation to healthcare, the practical impact of DL is becoming increasingly apparent.
However, DL faces significant security issues, particularly backdoor attacks.
Backdoor attacks typically embed a unique pattern (the backdoor trigger) into the training data, which establish a correlation between this pattern and a specific target label.
Consequently, a model trained on such data misclassifies inputs containing the trigger as the target label.
Researchers have proposed a range of backdoor attacks~\cite{dynamic,inputaware,invisible,wanet,sig,dfst,blend}, along with countermeasures aimed at detecting and mitigating backdoors in poisoned models~\cite{anp,nad,nc,neurotoxin,tabor,mntd,spectral_signature,spectre}.
However, without knowing the trigger pattern, it's challenging to accurately identify whether a model or dataset has been compromised, and the trigger pattern is typically not accessible until the attacker initiates the attack.

This paper focuses on backdoor mitigation~\cite{nad,anp,fine_pruning}. The goal is to remove the backdoor effect in a model such that trigger-inserted inputs cannot cause the target prediction. Backdoor mitigation usually assumes access to a few clean (usually $<5\%$) training samples without knowledge of the trigger pattern.
Existing backdoor mitigation techniques~\cite{nad,anp,fine_pruning,seam,nc,i-bau} are effective against prior attacks~\cite{badnet,trojannn,blend,sig,wanet,inputaware}.
However, they fall short in eliminating backdoor effects caused by advanced attacks~\cite{dfst, adaptive_blend}.
This is because these methods either retrain the entire model without precise guidance for reducing backdoor effects~\cite{nad,seam,nc} or directly prune some specific neurons~\cite{fine_pruning,anp}.
Such coarse-grained approaches fail to counter recent advanced attacks.
For instance, advanced attacks may hide backdoor behavior within benign neurons that primarily process normal features. In such cases, pruning these neurons would undesirably impact benign utility. On the other hand, retaining these neurons would preserve the backdoor behavior in the model.
To address the above challenge, we propose a novel backdoor mitigation method, \Tech{}.
It is based on the observation that, for various backdoored models, there exists a set of \textit{backdoor neurons}, responsible for backdoor behaviors. The activation values of these neurons for poisoned inputs are significantly higher than those for clean samples. Note that backdoor neurons may also play a role in benign feature extraction.
Given the absence of poisoned samples for accurately identifying backdoor neurons,, we propose to approximate a clean distribution on \textit{each individual} neuron using a small set of clean samples.
The approximation bounds the maximum activation value on each neuron.
During inference, our defense \Tech{} clips activations with a substantially large value to the approximated boundaries.
A straightforward idea is to apply a uniform percentile boundary, e.g., a threshold covering 98\% values, to bound the activation for all neurons. Our result in Figure~\ref{fig:naive_clip} (Section~\ref{sec:method}) reveals its limitation against advanced attacks, because it overlooks the fact that different neurons have various contributions.
While some neurons might be fully compromised, others could remain entirely benign.
To address this challenge, \Tech{} employs an optimization process that tailors a \textit{unique} boundary for each neuron.
The optimization is guided by a proxy accuracy measure on a small set of clean samples, serving as an approximation of the real accuracy on the test set. This is to precisely bound the accuracy degradation caused by the clipping.
This approximation is generally accurate, as evidenced by a ablation study detailed in Section~\ref{sec:appendix_abl_acc_deg}.
The process allows \Tech{} to meticulously tighten the boundaries to mitigate backdoor effects while ensuring the accuracy aligns with the defender's expectation.

Our main contributions are summarized as follows:
\begin{itemize}
    \item We introduce \Tech{} (``{\em A\underline{\sc U}tomated \underline{\sc N}eural D\underline{\sc I}stribution \underline{\sc T}ightening}''), an innovative backdoor mitigation method that approximates \textit{unique} distribution boundary for \textit{each} neuron, which is used to effectively dispel maliciously large activation caused by the backdoor.
    \item \Tech{} utilizes an optimization technique to dynamically refine and tighten unique boundaries for different neurons. This process is guided by the proxy accuracy on a few clean samples, which approximates the real test accuracy.
    \item Extensive experiments demonstrate \Tech{}'s effectiveness against 14 existing attacks, including 2 advanced attacks, outperforming 7 baseline defenses. Additionally, \Tech{} is generalizable to different datasets, network structures, and activation functions.
    We further show that \Tech{} is resilient to 3 adaptive attacks.
\end{itemize}

\smallskip \noindent
\textbf{Threat Model.}
Our threat model aligns with the existing literature~\cite{nad,anp,fine_pruning}, where the adversary provides a model that may potentially contain a backdoor to the user. The adversary holds the complete control over the training process and can deploy advanced attacks~\cite{dfst,adaptive_blend} to circumvent existing defenses. Prior to utilizing the model, the user applies defense techniques to mitigate any potential backdoor. The defender has access to a small portion (5\%) of the clean training data. She has no prior knowledge of the poisoned data.
The defense objective is to eliminate the backdoor effect without compromising the normal functionality, such as classification accuracy.

\section{Related Work} \label{sec:related}

\noindent
\textbf{Backdoor Attack.}
Recent literature has introduced a variety of backdoor attacks on image classification models.
Early works~\cite{badnet,trojannn} stamp static image patches on a small portion of training samples and mislabel them as the target class to poison the training dataset. Clean label attacks~\cite{clean_label,sleeper_agent,hidden_trigger} manipulate backdoor samples in feature space and leave their labels unchanged. Recently, more sophisticated transformations are utilized as backdoor triggers~\cite{blend,invisible,wanet,lotus,reflection,tao2022backdoor}. In addition, sample-specific backdoors generate different triggers for different inputs via generative models~\cite{inputaware,dynamic,dfst}, making them more stealthy and harder to detect.
Backdoor attacks can also be launched in a wide range of applications as well, such as natural language processing~\cite{badnl,hidden_killer,shen2022constrained}, self-supervised learning~\cite{badencoder,decree}, federated learning~\cite{backdoorfl,neurotoxin,flip} and even diffusion models~\cite{chou2023backdoor,elijah}.
In this paper, we focus on the image classification task.

\noindent
\textbf{Backdoor Defense.}
Various defenses have been proposed from multiple perspectives to safeguard AI models against backdoor attacks. Our approach falls under the category of \textit{Backdoor Mitigation}~\cite{fine_pruning,anp,nad,moth,i-bau,orthogonal,medic,tao2022deck}, which is widely acknowledged as a promising strategy. The primary goal is to cleanse the backdoor effect while retain the benign functionality of a given model.
Orthogonal to this, training-time defense~\cite{spectral_signature,spectre,anti_backdoor,decouple_backdoor,wang2022training,yan2023d} defenses distinguish between poisoned and clean samples based on their internal discrepancies/behaviors and sanitize the training set.
Trigger inversion~\cite{nc,abs,karm,tabor,beagle,dltnd,django,odscan} aims to detect whether a given model is poisoned or not via reverse-engineering the backdoor triggers.
Running time defenses~\cite{strip,februus} are designed to reject samples potentially carrying triggers during model inference.

\section{Limitation of Existing Backdoor Mitigation Methods} \label{sec:motivation}

Various methods have been proposed to address the challenge of mitigating the backdoor effects in poisoned models. They primarily fall into two categories: (1) Unlearning~\cite{nad,i-bau,nc,medic,seam,fst} and (2) Pruning~\cite{fine_pruning,anp,clp}.
Unlearning methods utilize training-based techniques, such as fine-tuning~\cite{fst}, distillation~\cite{nad}, and cloning~\cite{medic}, to eradicate the backdoor behaviors. These approaches are grounded in the catastrophe forgetting assumption~\cite{catastrophe}, positing that neural networks naturally tend to forget specific behaviors while continuously learning other patterns. For example, NAD~\cite{nad} employs standard fine-tuning to create a teacher model and then conveys only benign knowledge to the student model through knowledge distillation~\cite{knowledge_distill}.
In contrast, pruning methods involve the identification and removal of malicious neurons. They speculate that there exists a small subset of neurons responsible for backdoor behaviors, and the removal of these neurons eliminates the backdoor impact.
For instance, ANP~\cite{anp} identifies malicious neurons based on sensitivity analysis on clean samples and effectively prunes them.
In the following, we delve into the limitations inherent to both unlearning and pruning methods and introduce our idea to address the challenges.

\begin{figure}[t]
    \centering
    \includegraphics[width=1\textwidth]{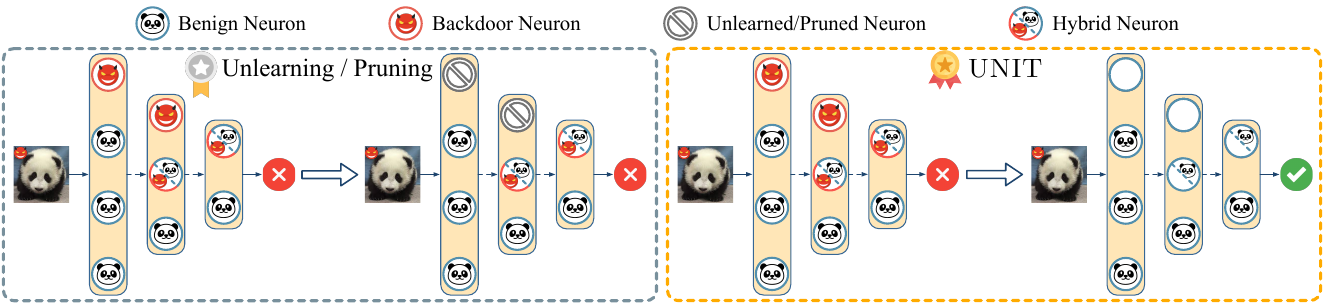}
    \caption{Limitation of existing backdoor mitigation methods}
    \label{fig:limitation}
\end{figure}

\smallskip \noindent
\textbf{Coarse-grained Repair.}
Recent advanced attacks~\cite{dfst,adaptive_blend} manage to conceal backdoor behavior within benign neurons, creating hybrid neurons that withstand existing mitigation methods. A prevalent limitation in current techniques lies in their coarse-grained nature, which is inadequate against advanced attacks. Essentially, these methods struggle to operate inside individual neurons to eliminate the backdoor component while preserving the benign portion.
In addition, an implicit requirement in backdoor mitigation is the preservation of benign functionality. 
In other words, the benign accuracy of the repaired model should not suffer significant degradation. 
This constraint limits the efficacy of both unlearning and pruning methods.
For example, pruning may either remove or leave an entire neuron untouched. When dealing with hybrid neurons, directly pruning them would significantly diminish clean classification performance. Conversely, retaining such neurons would maintain the backdoor behaviors.
Figure~\ref{fig:limitation} conceptually illustrates such limitation of existing methods. The left dashed box shows the mitigation of an advanced attack~\cite{dfst,adaptive_blend}. The left half presents the process of a poisoned image (depicted as a panda with a red trigger at the top-left) in a backdoored model. Notably, the model comprises three types of neurons: (1) Benign neurons (depicted as cartoon pandas) primarily extracting benign features, (2) Backdoor neurons (depicted as red devils) processing backdoor behaviors, and (3) Hybrid neurons (depicted as half panda and half devil) serving both purposes. Following the model inference, the output corresponds to the misclassified attack target label, indicated by a red cross.
The right half of the left figure portrays the model after repair through unlearning and pruning.
Observe that backdoor neurons are effectively unlearned or pruned, whereas hybrid neurons, which exhibit both benign and malicious behaviors, remain unaffected. This is because eliminating these hybrid neurons could lead to a substantial decrease in accuracy for benign tasks. However, the presence of these hybrid neurons can still contribute to the persistence of a high attack success rate due to their involvement in backdoor behaviors.
This highlights the limitations of current mitigation techniques.

\smallskip \noindent
\textbf{Heavily Dependent on Meticulous Parameter Tuning.}
Existing approaches heavily rely on meticulous parameter tuning to achieve optimal performance against various attacks. For instance, pruning techniques demand a careful determination of the pruning rate, adjusted on a case-by-case basis. The extent of neuron removal directly influences the model's overall accuracy; excessive pruning can deteriorate performance, while insufficient pruning may not adequately counteract the backdoor effect.
Our empirical analysis, detailed in Appendix~\ref{sec:appendix_param}, highlights the pronounced sensitivity of the existing methods to parameter adjustments.
This sensitivity presents a notable limitation, undermining the generalizability and practical applicability of these methods.

\smallskip \noindent
\textbf{Our Idea:} \textit{Automated Neural Distribution Tightening.}
We introduce a novel technique \Tech{}, which automatically approximates and tightens a unique distribution boundary for each neural activation. Subsequently during inference, it clips activation values that exceed the boundary, targeting potential backdoor activation.
\Tech{} employs an optimization based method to automatically refine the activation boundary for individual neurons. It is guided by a proxy accuracy measured on a small set (<5\%) of clean samples, which approximates the real test accuracy. This approximation is generally accurate, as evidenced by a ablation study detailed in Section~\ref{sec:appendix_abl_acc_deg}.
The process involves a dynamic adjustment of boundaries: if the observed proxy accuracy degradation is below the defender's expectation, the boundary is further tightened. Conversely, 
the boundary is relaxed to restore the accuracy. This ensures a balanced approach to maintaining benign accuracy while eliminating backdoors.
\Tech{} operates with a high degree of granularity, analyzing and adjusting unique boundaries for individual neurons.
The right figure in Figure~\ref{fig:limitation} visualizes positive outcomes achieved through \Tech{}. Notably, both backdoor neurons and the backdoor portion of hybrid neurons are deactivated.

Moreover, compared with existing methods, \Tech{} is an automated technique that does not require \textit{meticulous} parameter tuning. The defender is only required to specify a bound of accuracy degradation to balance benign accuracy and backdoor mitigation. The \textit{parameter-efficient} characteristic of \Tech{} emphasizes the generalizability and practicality of \Tech{}.

\section{Design of \Tech{}} \label{sec:method}

\subsection{Notations} \label{sec:notation}
We provide formal notations of deep neural network operations before delving into our methodology.
Following existing works~\cite{clp,preact}, we consider a typical neural network for a classification task of $C$ classes. The dataset $\mathcal{D}$ is composed of numerous pairs $(x,y) \sim \mathcal{D}$, where each sample $x \in \R^d$ and its corresponding label $y \in \{1, 2, \cdots C\}$. 
The input dimension $d$ can be complex, e.g., $d = d_c \times d_w \times d_h$ for RGB images, where $d_c$, $d_w$, and $d_h$ represent the number of channels, width, and height, respectively.
The training objective is to derive a classifier $M: \R^d \rightarrow \{1, 2, \cdots, C\}$.
Consider a deep neural network consisting of $L$ layers:
\begin{equation} \label{eq:cls}
    M = g \circ \phi \circ f^L \circ \cdots \circ \phi \circ f^l \circ \cdots \circ \phi \circ f^{1},
\end{equation}
\noindent where $f^l$ denotes the feature extraction function at $l$-th layer ($1 \leq l \leq L$), $\phi$ represents the non-linear activation function, e.g., ReLU~\cite{relu}, and $g$ is the fully connected layer following the extraction layers, responsible for aggregating features for class prediction.

\smallskip \noindent \textbf{Neural Activation.}
To analyze the internal statistics of the model, we further define the sub-network that terminates at the $l$-th activation layer as $F^l$:
\begin{equation}
    F^{l} = \phi \circ f^l \circ \cdots \circ \phi \circ f^{1}
\end{equation}
\noindent Therefore, given an input sample $x$, its activation value at $l$-th layer is $F^{l}(x)$.
This activation value is typically multi-dimensional.
If the $l$-th layer consists of $K$ neurons, the \textit{neural activation} of the $k$-th neuron in this layer is denoted as $F^{l}_{k}(x)$.

\begin{figure}[t]
    \centering
    \includegraphics[width=1\textwidth]{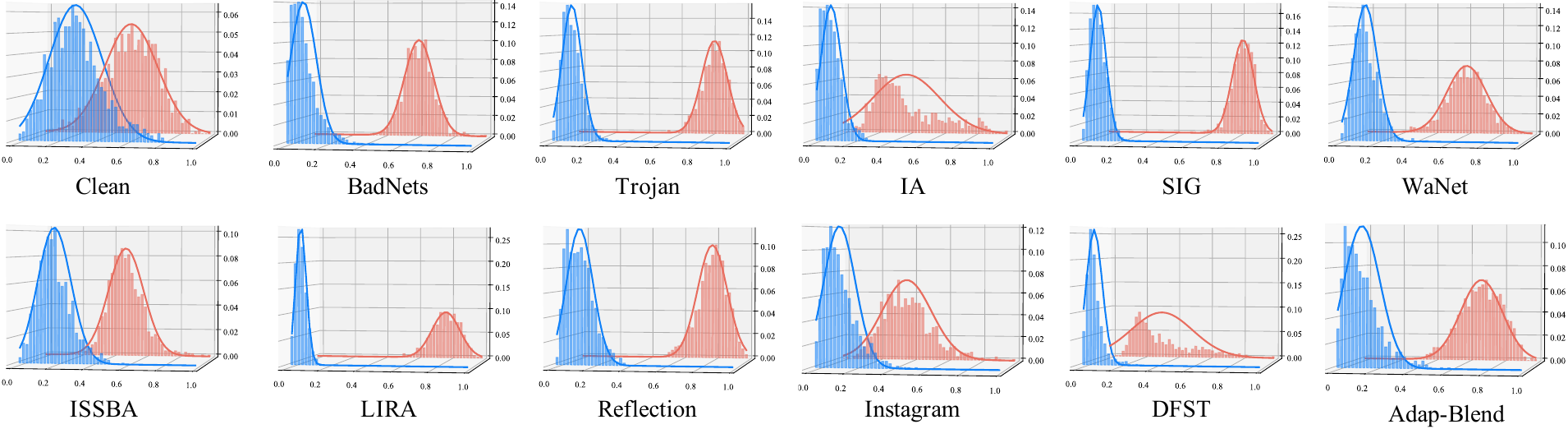}
    \caption{Neural activation distribution for benign and poisoned samples}
    \label{fig:observe}
\end{figure}

\subsection{Key Observations of Neural Activation}
The backdoor behavior can be activated by the trigger on backdoored models. To illustrate how such input pattern flips the output prediction, we delve into the model internals, particularly examining the neural activation values of both clean and poisoned samples.
We use the CIFAR-10 dataset and ResNet18 architecture as our subject and visualize the neural activation distribution of a clean model and a range of backdoored models by various attacks, including BadNets~\cite{badnet}, Trojan~\cite{trojannn}, IA~\cite{inputaware}, SIG~\cite{sig}, WaNet~\cite{wanet}, ISSBA~\cite{invisible}, LIRA~\cite{lira}, Reflection~\cite{reflection}, Instagram~\cite{trojannn}, DFST~\cite{dfst}, and Adap-Blend~\cite{adaptive_blend}.
To gain insights into the influence of poisoned samples on model behavior, we utilize Shap~\cite{shap}, a deep learning interpreter, to identify 1\% of the most important neurons in the 12th layer of each model when processing poisoned samples. These selected neurons, designated as \textit{backdoor neurons}, are responsible for the backdoor behavior. Subsequently, our analysis involves comparing the activation values of these neurons across 1,000 clean and 1,000 poisoned samples.
It's worth noting that as there is no predefined trigger for the clean model, we employ the BadNets trigger to generate dummy poisoned samples for analysis.
By applying PCA~\cite{pca} for dimensionality reduction, we visualize the neural activation distributions in Figure~\ref{fig:observe}. The blue plots represent the activation distributions of clean inputs, while the red plots depict the distributions of poisoned samples.
Observe that the neural activation distributions of clean and poisoned samples are indistinguishable in the clean model. Conversely, in models subjected to backdoor attacks, it is evident that there exists a large distribution shift between clean and poisoned samples.
Notably, the neural activation values for poisoned inputs are significantly greater than those for clean inputs.
This disparity underscores that backdoor triggers significantly change the neural activation distribution for specific \textit{backdoor neurons}, subsequently leading to the target misclassification.

\smallskip \noindent \textbf{Distinguished Fine-grained Observation.}
Existing papers~\cite{ac,spectral_signature,adaptive_blend} have observed the latent separability between clean and poisoned samples, primarily focusing on the features of the \textit{entire layer}. Nonetheless, recent advanced attacks~\cite{adaptive_blend} and the adaptive attacks detailed in Section~\ref{sec:appendix_adaptive} manage to diminish this \textit{layer-level} feature distinction. However, these approaches fall short in eliminating separability at the \textit{neural activation} level, as shown in Figure~\ref{fig:observe}. This highlights a clear distinction between our fine-grained observation and existing literature.

\subsection{Overview of \Tech{}}
\begin{figure}[t]
    \centering
    \begin{minipage}[c]{0.48\textwidth}
        \centering
        \includegraphics[width=1\textwidth]{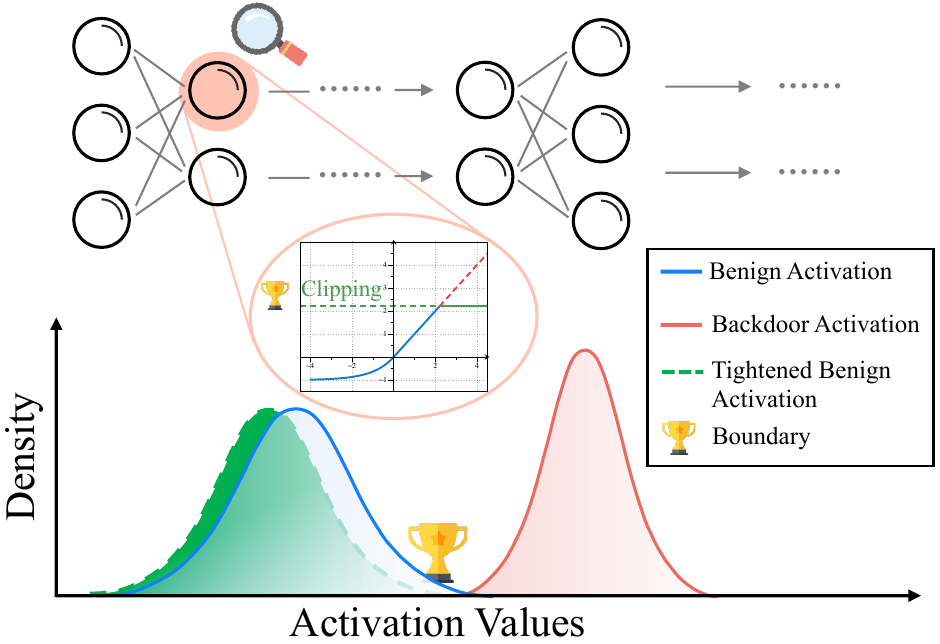}
        \caption{Overview of \Tech{}}
        \label{fig:overview}
    \end{minipage}
    \hfill
    \begin{minipage}[c]{0.48\textwidth}
        \centering
        \includegraphics[width=1\textwidth]{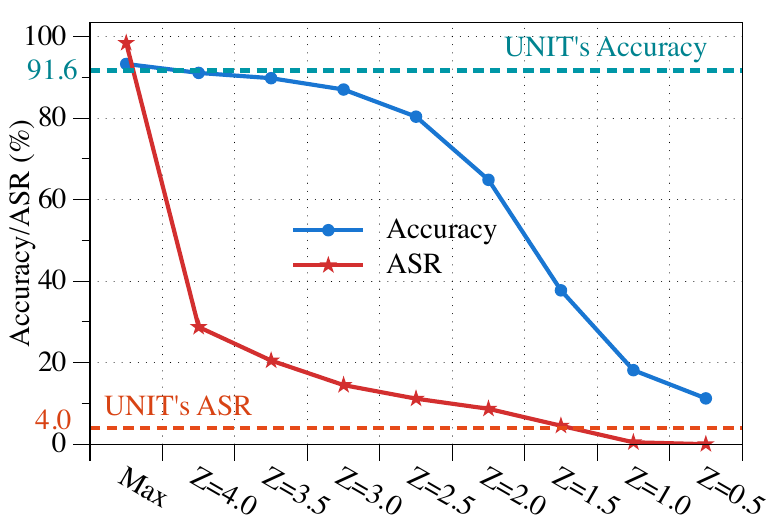}
        \caption{Limitation of straightforward clipping}
        \label{fig:naive_clip}
    \end{minipage}
\end{figure}

Our observation reveals a substantial increase in neural activation compared to benign ones on backdoor neurons given poisoned samples. Building upon this insight, we introduce \Tech{}, a novel approach that approximates a tight benign distribution for each neuron based on a small subset of clean training data. \Tech{} then strategically clips activation values that surpass the distribution boundary.
The necessity for this approximation stems from the unavailability of poisoned samples in typical scenarios. Hence, it is challenging to precisely identify the backdoor neurons. To deal with the problem, \Tech{} applies its approximation across all neurons, including both benign and backdoor ones. Furthermore, we refine the approximated benign distribution to be as tight as possible, aiming to effectively mitigate the backdoor behavior.
The overview of \Tech{} is depicted in Figure~\ref{fig:overview}, using a typical neuron as an example. The x-axis represents neural activation values for different samples, while the y-axis denotes sample density corresponding to these values. In this depiction, the benign neural activation is shown in blue, and the poisoned neural activation is in red.
\Tech{} approximates a tight distribution based on a few clean samples, as illustrated in the green region.
To mitigate the backdoor effect during model inference, \Tech{} constrains the neural activation values by clipping those that exceed the approximated boundary. In the zoomed-in plot, rather than allowing activation values to extend along the red lines (which represent maliciously large values), \Tech{} ensures that values remain within the green boundary.
While this strategy might entail a minor compromise in accuracy for clean samples, it is remarkably effective in neutralizing the backdoor effects of poisoned samples, thereby enhancing the model's security and integrity.

\begin{algorithm}[h]
   \caption{Automated Neural Distribution Tightening}
   \label{alg:clip}
\begin{algorithmic}[1]
   \State {\bfseries Input:} Subject model $M$, Accuracy drop expectation $\epsilon$, Training data $\{(x_i^t, y_i^t)\}_{i=1}^{n_t}$, Validation data $\{(x_i^v, y_i^v)\}_{i=1}^{n_v}$, Initial benign distribution boundary $\sigma_0$, Initial trade-off coefficient $\alpha_0$, Optimization steps $S$, and Learning rate $\eta$.
   \State {\bfseries Initialize:} $\sigma=\sigma^{\star}=\sigma_0$, $\alpha=\alpha_0$
   \LineComment[0\dimexpr\algorithmicindent]{{\small \textit{Calculate original accuracy on validation samples}}}
   \State $P_0 = \frac{1}{n_v} \sum_{i=1}^{n_v}\mathbbm{1}({M(x^v_i) = y^v_i})$
   \For{$s=1$ {\bfseries to} $S$}
   \LineComment[1\dimexpr\algorithmicindent]{{\small \textit{Cross-entropy loss plus boundary penalty}}}
   \State $\mathcal{L} = \frac{1}{n_t} \sum_{i=1}^{n_t} \mathcal{L}_{CE}(M_{\sigma}(x_i^t), y_i^t) + \alpha \cdot ||\sigma||_{1}$
   \State $\sigma = \sigma - \eta \cdot \frac{\partial \mathcal{L}}{\partial \sigma}$
   \LineComment[1\dimexpr\algorithmicindent]{{\small \textit{Calculate accuracy when applying current bound}}}
   \State $P^{\prime} = \frac{1}{n_v} \sum_{i=1}^{n_v}\mathbbm{1}({M_{\sigma}(x^v_i) = y^v_i})$
   \If{$P_0 - P^{\prime} > \epsilon$}
   \State $\alpha = \alpha / 2$
   \Else
   \State $\alpha = \alpha \cdot 2$
   \EndIf
   \LineComment[1\dimexpr\algorithmicindent]{{\small \textit{Update the best boundary value}}}
   \If{$P^{\prime} \geq P_0 - \epsilon$ and $||\sigma||_1 < ||\sigma^{\star}||_1$}
   \State $\sigma^{\star} = \sigma$
   \EndIf
   \EndFor
   \State {\bfseries Return:} $\sigma^{\star}$
\end{algorithmic}
\end{algorithm}

\subsection{Design Details}
In this section, we formally present the design of \Tech{}. Specifically, we detail the process of automatically tightening the neural distribution based on a small portion of clean training samples. The goal is to effectively eliminate maliciously high neural activation, which represents the backdoor behavior.

\smallskip \noindent
\textbf{Objective.}
Following the notation of neural activation in Section~\ref{sec:notation}, we formally define the objective of \Tech{}.
For any input $x$ and its neural activation at $l$-th layer and $k$-th neuron $F^l_k(x)$, \Tech{} derives an upper bound value $\sigma^l_k$ such that its neuron activation is bounded as
\begin{equation}
    \hat{F}^l_k(x) = b_{\sigma^l_k}(F^l_k(x)) = 
    \begin{cases}        
        F^l_k(x) & \text{if } F^l_k(x) \leq \sigma^l_k, \\
        \sigma^l_k & \text{otherwise.}
    \end{cases}
\end{equation}
\noindent Note that $\sigma^l_k$ can be a feature map when $F^l$ is a convolution layer.
Then the classifier defined in Equation~\ref{eq:cls} can be reformatted as:
\begin{equation}
    M_{\sigma} = g \circ b_{\sigma^L} \circ \phi \circ f^L \circ \cdots \circ b_{\sigma^l} \circ \phi \circ f^l \circ \cdots \circ b_{\sigma^1} \circ \phi \circ f^{1},
\end{equation}
\noindent where $\sigma^l$ denotes the bounding value at the $l$-th layer. Suppose there are $K$ neurons at this layer, then $\sigma^l = \{\sigma^l_1,\sigma^l_2,\cdots,\sigma^l_K\}$. Similarly, $\sigma = \{\sigma^1,\sigma^2,\cdots,\sigma^L\}$.
The objective of \Tech{} is to mitigate the backdoor effects while preserve the benign utility. Therefore, for any input $x$ of class $y$ and its poisoned version $x \oplus T$ with the attack target label $y_T$, where $T$ denotes the backdoor trigger,
\begin{equation}
    M_{\sigma}(x) = y \text{,\quad} M_{\sigma}(x \oplus T) \neq y_T.
\end{equation}

A straightforward idea is to employ a uniform percentile threshold for all neural activation values.
However, it can be inaccurate and coarse-grained as different neurons vary in their contributions to backdoor effects.
Figure~\ref{fig:naive_clip} demonstrates the effectiveness of this approach against the DFST~\cite{dfst} attack (launched using CIFAR-10 and ResNet-18), where the original model achieves a clean accuracy of 92.25\% and an ASR of 99.77\%.
The x-axis represents various uniform clipping percentiles while the y-axis shows the corresponding accuracy and ASR after clipping.
"Max" indicates setting the boundary at each neuron's maximum activation value. In other cases, we assume a Gaussian distribution of the activation and employ the Z-score for percentile approximation. For example, "Z=3.0" signifies setting the boundary at the mean activation value plus three times its standard deviation, aligning with the 0.98 percentile.
We can observe that even with a moderate clean accuracy of 90\% (Z=3.5), the ASR remains notably high at 20\%. Conversely, reducing the ASR to 4\% (Z=1.5) leads to a drastic decrease in accuracy, down to 40\%. This highlights the method's limitation against advanced attacks.

Our approach, on the other hand, utilizes an optimization-based technique to meticulously approximate and tighten a unique boundary for each individual neuron, which outperforms the straightforward approach as illustrate in the blue and red dashed lines in Figure~\ref{fig:naive_clip}. Note that \Tech{} is able to reduce the ASR to 4.0\% while maintain a high accuracy as 91.6\%.
The details of \Tech{} are outlined in Algorithm~\ref{alg:clip}, which comprises two main stages: (1) Initialization (Line 1-3), where clean training samples are gathered to approximate a loose benign boundary, and (2) Automated Tightening (Line 4-16), dedicated to refining the approximated boundary with the guidance of clean accuracy.

\smallskip \noindent
\textbf{Initialization.}
Lines 1-3 present the initialization stage, where input variables are defined, with $M$ representing the model for defense, and $\epsilon$ indicating the customized accuracy drop expectation (defaulted to $2\%$). Following the threat model in Section~\ref{sec:intro}, the defender has access to a small set of clean training samples for the defense process. The data is further split into training samples $\{(x_i^t, y_i^t)\}_{i=1}^{n_t}$ and validation samples $\{(x_i^v, y_i^v)\}_{i=1}^{n_v}$, where $n_t$ and $n_v$ denote the number of training and validation samples, respectively.
Typically, the ratio $\frac{n_v}{n_t}$ is set to $\frac{1}{4}$. The split training samples are used for optimization, while validation samples guide the tightening strength.
A loose distribution for clean samples is approximated, initializing the distribution boundary of each neuron as $\sigma_0$. This initial boundary is set as the mean activation value over the training sample plus four times the standard deviation (Z-score=4 in the straightforward approach).
The initial trade-off coefficient between benign accuracy and the tightened distribution boundary is denoted as $\alpha_0$, with a default value set to 0.001. This value signifies that the tightening process starts with low strength.
Additionally, $S$ represents the number of optimization steps, and $\eta$ denotes the learning rate.
Typically, 50 steps prove sufficient to approximate a suitably tight boundary.
For optimization, we utilize the Adam optimizer with a learning rate set to $\eta=0.001$, a standard configuration.
Line 2 initializes the optimized distribution boundary $\sigma$, optimal boundary $\sigma^{\star}$, and trade-off coefficient $\alpha_0$ with their default values. In Line 3 calculates the initial accuracy ($P_0$) of model $M$ on validation samples, where $\mathbbm{1}({M(x^v_i) = y^v_i})$ denotes the number of samples which are correctly classified by $M$.

\smallskip \noindent
\textbf{Automated Tightening.}
Lines 4-16 outline the optimization procedure for tightening the benign distribution. In each optimization step, the goal is to tighten the boundary while maintaining benign accuracy within the specified expectation $\epsilon$.
Line 5 calculates the loss, consisting of two terms: the cross-entropy loss on the training samples and the penalty on boundary scale. We use L-1 norm of $\sigma$ to measure the tightness of the current boundary. A small value of $||\sigma||_1$ means a tight boundary.
The trade-off between these two loss terms is controlled by $\alpha$. The boundary $\sigma$ is optimized using gradient descent in Line 6.
Lines 7-12 dynamically adjust the trade-off value $\alpha$ based on the accuracy on validation samples. In Line 7, the current accuracy $P^{\prime}$ on validation samples is calculated given the optimized $\sigma$. If the accuracy drop $P_0 - P^{\prime}$ exceeds the expectation $\epsilon$ (Line 8), the trade-off coefficient $\alpha$ is reduced by half (Line 9), prioritizing the restoration of benign accuracy. Otherwise, $\alpha$ is increased twice to further tighten the benign distribution boundary (Line 11).
Lines 13 to 15 update the optimal boundary $\sigma^{\star}$ if it maintains benign accuracy while being more tightened.
Finally, Line 17 returns the optimal boundary $\sigma^{\star}$ and \Tech{} applies the optimal boundary to the model ($M_{\sigma^{\star}}$) during inference.

\section{Evaluation} \label{sec:eval}
In this section, we comprehensively evaluate the performance of \Tech{} across diverse scenarios.
In Section~\ref{sec:effective}, we assess the effectiveness of \Tech{} by comparing it against 7 state-of-the-art backdoor mitigation baselines across 14 types of backdoor attacks.
In addition, we demonstrate the generalizability of \Tech{} by the evaluation on four datasets and six network architectures.
We assess the time cost of \Tech{} in Section~\ref{sec:efficient} and study the effect of \Tech{} on clean models in Section~\ref{sec:clean_effect}.
In Section~\ref{sec:additional_exp}, we present additional evaluations of \Tech{} against the latest backdoor attacks and comparisons with recent baselines. We also include a series of evaluations on adaptive attacks and ablation studies.

\subsection{Experiment Setup}

\smallskip \noindent
\textbf{Baselines and Settings}
We employ 14 backdoor attacks, (1) BadNets~\cite{badnet}, (2) Trojan~\cite{trojannn}, (3) CL~\cite{clean_label} (4) Dynamic backdoor~\cite{dynamic}, (5) IA~\cite{inputaware}, (6) Reflection~\cite{reflection}, (7) SIG~\cite{sig}, (8) Blend~\cite{blend}, (9) WaNet~\cite{wanet}, (10) ISSBA~\cite{invisible}, (11) LIRA~\cite{lira}, (12) Instagram filter~\cite{trojannn}, (13) DFST~\cite{dfst}, and (14) Adap-Blend~\cite{adaptive_blend}.
We use the default configuration following the original papers, such as trigger patterns, sizes, poisoning strategies, etc.
We compare \Tech{} with 7 state-of-the-art backdoor mitigation methods, (1) standard fine-tuning (FT), (2) FP~\cite{fine_pruning}, (3) NAD~\cite{nad}, and (4) ANP~\cite{anp}, (5) NC~\cite{nc}, (6) I-BAU~\cite{i-bau} and (7) SEAM~\cite{seam}.
We follow the configuration in the original papers to conduct experiments.
All the methods have access to the same amount of training data, e.g., 5\%.
Details of backdoor attack and defense baselines can be found in Appendix~\ref{sec:appendix_details}.
For \Tech{}, we set the expected accuracy degradation as 2\%.

\smallskip \noindent
\textbf{Evaluation Metrics.}
We use two metrics: (1) clean accuracy (Acc.), and (2) attack success rate (ASR). Clean accuracy measures the normal functionality of the subject model on classifying clean inputs. ASR measures the backdoor effect, which is the ratio of poisoned samples correctly misclassified to the target label.
A good defense shall reduce the ASR while preserving the clean accuracy.

\begin{table}[t]
    \centering
    \tiny
    \tabcolsep=0.73pt
    \caption{Comparison of \Tech{} with 7 backdoor mitigation baselines against 14 backdoor attacks. Results are measured in percentages (\%). All methods have access to 5\% of the clean training data. The best results are highlighted in bold.}
    \label{tab:baseline}
    \begin{tabular}{lgrgrgrgrgrgrgrgrgr}
         \toprule
         \multirow{2}{*}{\textbf{Attacks}} & \multicolumn{2}{c}{\textbf{Original}} & \multicolumn{2}{c}{\textbf{FT}} & \multicolumn{2}{c}{\textbf{FP}} & \multicolumn{2}{c}{\textbf{NAD}} & \multicolumn{2}{c}{\textbf{ANP}} & \multicolumn{2}{c}{\textbf{NC}} & \multicolumn{2}{c}{\textbf{I-BAU}} & \multicolumn{2}{c}{\textbf{SEAM}} & \multicolumn{2}{c}{\textbf{\Tech{}}} \\
         \cmidrule(lr){2-3} \cmidrule(lr){4-5} \cmidrule(lr){6-7} \cmidrule(lr){8-9} \cmidrule(lr){10-11} \cmidrule(lr){12-13} \cmidrule(lr){14-15} \cmidrule(lr){16-17} \cmidrule(lr){18-19}
         ~ & \cellcolor{white}{Acc.} & \cellcolor{white}{ASR} & \cellcolor{white}{Acc.} & \cellcolor{white}{ASR} & \cellcolor{white}{Acc.} & \cellcolor{white}{ASR} & \cellcolor{white}{Acc.} & \cellcolor{white}{ASR} & \cellcolor{white}{Acc.} & \cellcolor{white}{ASR} & \cellcolor{white}{Acc.} & \cellcolor{white}{ASR} & \cellcolor{white}{Acc.} & \cellcolor{white}{ASR} & \cellcolor{white}{Acc.} & \cellcolor{white}{ASR} & \cellcolor{white}{Acc.} & \cellcolor{white}{ASR} \\
         \midrule
BadNets                  & 94.82         & 100.0         & 90.91      & 9.78      & 89.68      & 3.52      & 92.41      & 4.79       & 91.35      & 3.26       & \textbf{93.04}      & \textbf{0.34}      & 91.60       & 3.66        & 91.61       & 1.05       & 92.48          & 0.78           \\
Trojan                   & 94.73         & 100.0         & 91.63      & 35.11     & 90.76      & 31.14     & 91.52      & 22.30      & 92.37      & 58.88      & 91.89      & 4.01      & 90.73       & 11.58       & 92.28       & 12.69      & \textbf{92.38} & \textbf{2.17}  \\
CL                       & 94.58         & 98.46          & 90.34      & 58.72     & 87.71      & 3.69      & 88.47      & 4.42       & 89.92      & 18.18      & 90.72      & 1.79      & 88.75       & 5.52        & 92.02       & 23.04      & \textbf{92.21} & \textbf{1.09}  \\
Dynamic                  & 95.08         & 100.0         & 89.11      & 9.29      & 84.93      & 3.23      & 89.26      & 2.34       & 91.99      & 3.09       & 92.09      & 1.78      & 92.48       & 1.63        & 92.61       & 3.22       & \textbf{92.77} & \textbf{1.54}  \\
IA                       & 91.15         & 97.96          & 88.44      & 2.92      & 89.71      & 82.20     & 88.51      & 2.67       & 89.05      & 5.49       & 89.32      & 1.12      & 89.79       & 62.45       & 89.77       & 1.23       & \textbf{89.93} & \textbf{1.03}  \\
Reflection               & 93.29         & 99.33          & 91.38      & 74.77     & 89.68      & 84.51     & 90.99      & 52.97      & 90.66      & 93.28      & 91.38      & 93.31     & 89.94       & 87.85       & 90.54       & 21.37      & \textbf{91.44} & \textbf{6.63}  \\
SIG                      & 94.97         & 99.80          & 91.29      & 63.94     & 90.88      & 1.03      & 91.69      & 10.46      & 90.80      & 36.79      & 91.70      & 97.88     & 91.51       & 22.11       & \textbf{92.57}       & \textbf{0.68}       & 92.48          & 1.74           \\
Blend                    & 94.62         & 100.0         & 90.68      & 7.30      & 91.47      & 2.01      & 91.62      & 3.32       & 91.04      & 16.79      & 91.90      & 1.53      & 91.43       & 3.61        & 91.38       & 1.80       & \textbf{91.99} & \textbf{1.18}  \\
WaNet                    & 94.36         & 99.80          & 90.32      & 2.85      & 91.48      & 1.48      & 92.36      & 1.91       & \textbf{91.99}      & \textbf{0.61}       & 90.60      & 0.97      & 89.67       & 12.01       & 91.34       & 1.44       & 91.02          & 2.44           \\
ISSBA                    & 94.55         & 100.0         & 91.40      & 4.17      & 90.79      & 2.11      & 92.45      & 2.43       & 92.42      & 2.98       & \textbf{92.52}      & \textbf{0.46}      & 83.03       & 84.58       & 91.17       & 3.00       & 91.84          & 1.57           \\
LIRA                     & 95.11         & 100.0         & 91.42      & 15.09     & 89.58      & 14.76     & 91.64      & 2.06       & 91.98      & 47.91      & 92.11      & 1.17      & 92.18       & 12.65       & 92.18       & 3.02       & \textbf{92.29} & \textbf{0.58}  \\
Instagram                & 94.62         & 99.59          & 91.40      & 29.25     & 90.38      & 8.03      & 89.50      & 7.17       & 90.10      & 5.10       & 90.19      & 15.88     & 89.25       & 7.24        & 91.35       & 5.89       & \textbf{91.43}          & \textbf{4.98}           \\
DFST                     & 93.25         & 99.77          & 90.88      & 35.22     & 90.66      & 14.03     & 91.05      & 14.59      & 89.70      & 20.51      & 91.22      & 24.77     & 89.12       & 6.19        & 91.22       & 12.93      & \textbf{91.64}          & \textbf{4.02}           \\
Adap-Bl.               & 94.22         & 82.80          & 90.15      & 48.76     & 87.62      & 31.36     & 90.42      & 49.50      & 90.80      & 69.51      & 90.33      & 18.25     & 90.81       & 19.97       & 89.58       & 24.19      & \textbf{90.84} & \textbf{15.03} \\
\midrule
\textbf{Average}                  & 94.26         & 98.39          & 90.57      & 28.37     & 89.67      & 20.22     & 90.85      & 12.92      & 91.01      & 27.31      & 91.36      & 18.80     & 90.02       & 24.36       & 91.48       & 8.08      & \textbf{91.77} & \textbf{3.20}       
          \\
         \bottomrule
    \end{tabular}
\end{table}

\subsection{Effectiveness of \Tech{}} \label{sec:effective}

\subsubsection{Comparison with Existing Baselines}
We conducted a comprehensive evaluation of \Tech{} by comparing it with 9 baseline methods across 14 distinct backdoor attacks on the CIFAR-10 dataset using the ResNet18 architecture for evaluation, with all defenses having access to 5\% of the training set.
Table~\ref{tab:baseline} summarizes the results. The first column enumerates different backdoor attacks, while the ``No Defense'' column displays the original performance of backdoored models. The subsequent columns detail the performance of models repaired by various defenses, with ``Acc.'' denoting clean test accuracy and ``ASR'' representing the attack success rate of backdoor attacks.
Notably, \Tech{} consistently outperforms others in reducing ASR and maintaining high clean accuracy. In instances such as Reflection, DFST, and Adap-Blend attacks, existing defense methods struggle to eliminate the backdoor effect, often retaining over 20\% ASR. The sophistication of these attacks, characterized by larger triggers and specialized poisoning strategies, poses a challenge to conventional defenses.
For instance, the state-of-the-art Adap-Blend attack relaxes the latent separability assumption and utilizes asymmetric triggers to enhance backdoor resilience.
Despite the complexity, \Tech{} reduces the ASR to less than 7\% for Reflection and DFST, outperforming existing methods, even mitigating the Adap-Blend attack to 15\%.
However, it's worth noting that \Tech{} doesn't outperform baselines in all scenarios.
For instance, ANP performs better on WaNet than \Tech{}.
This is due to the pervasive and sample-specific triggers of WaNet attacks. 
They resemble natural features and hence make the poisoned activation distribution less distinguishable from the clean one.
Despite this, \Tech{} still demonstrates effectiveness by mitigating the backdoor effect to an ASR of under 2.5\%.
Furthermore, we evaluate \Tech{} on two latest backdoor attacks, i.e., NARCISSUS~\cite{narcissus} and COMBAT~\cite{combat}, and compare the performance with five state-of-the-art baselines, i.e., CLP~\cite{clp}, FST~\cite{fst}, RNP~\cite{rnp}, FT-SAM~\cite{ft-sam} and Super-FT~\cite{super-ft}. The results in Appendix~\ref{sec:appendix_baseline} demonstrate \Tech{}'s superior performance over these methods.

\begin{figure*}[t]
    \centering
    \begin{minipage}{0.48\textwidth}
        \includegraphics[width=1\textwidth]{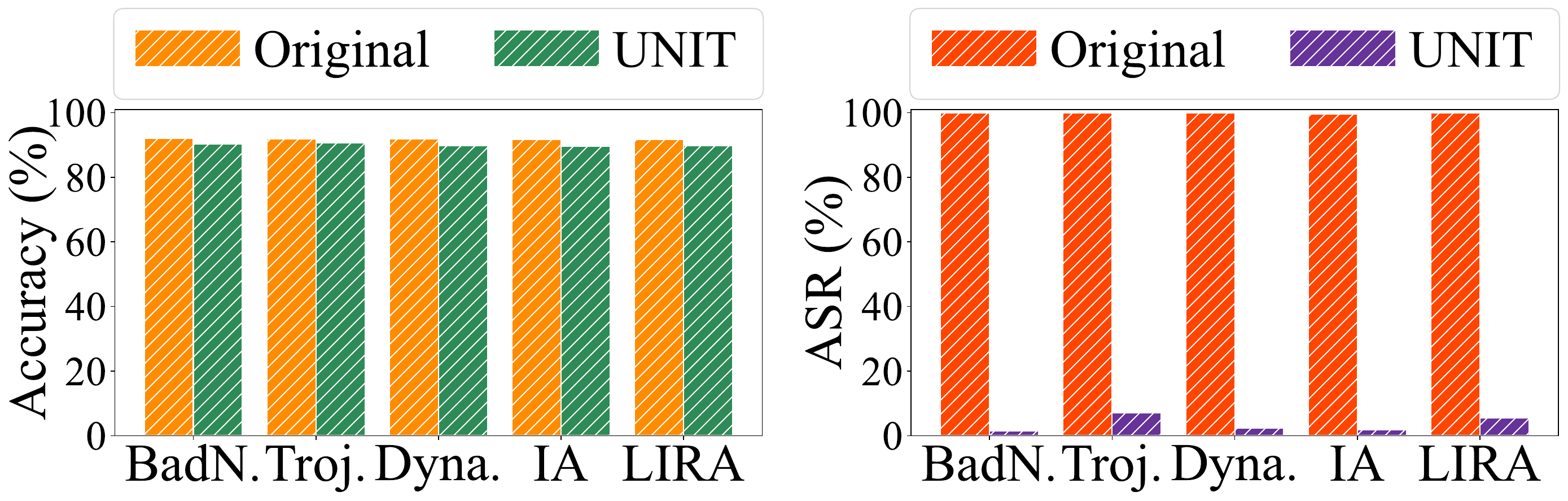}
        \centering (a) CIFAR-10 and VGG11
    \end{minipage}
    \begin{minipage}{0.48\textwidth}
        \includegraphics[width=1\textwidth]{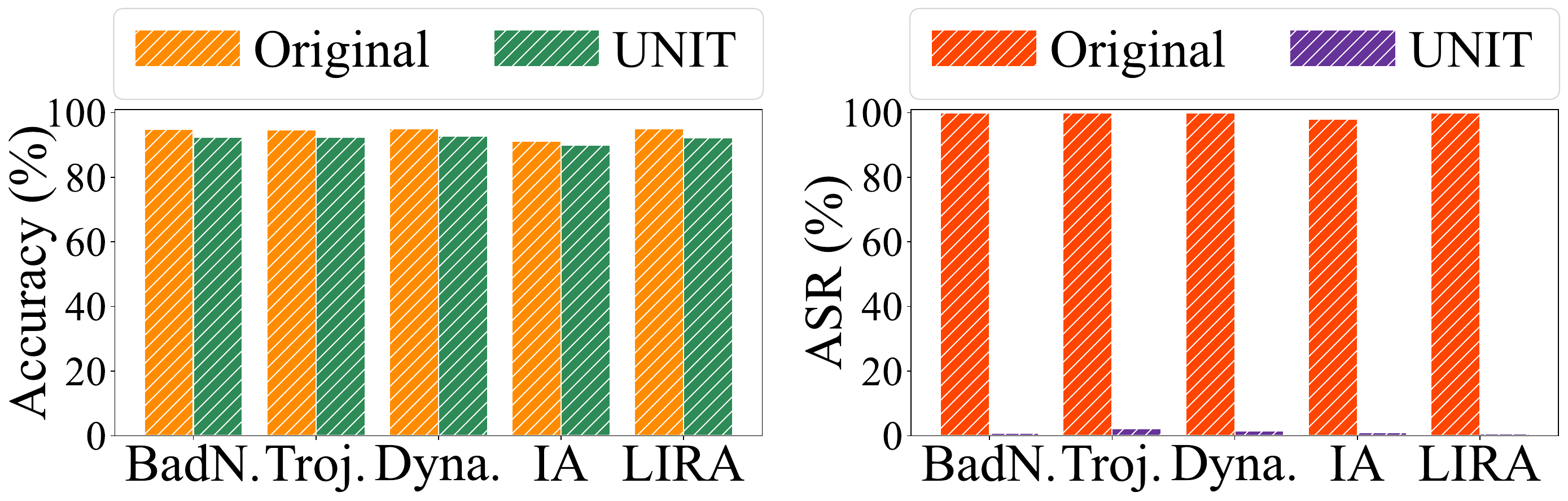}
        \centering (b) CIFAR-10 and ResNet18
    \end{minipage}
    \begin{minipage}{0.48\textwidth}
        \includegraphics[width=1\textwidth]{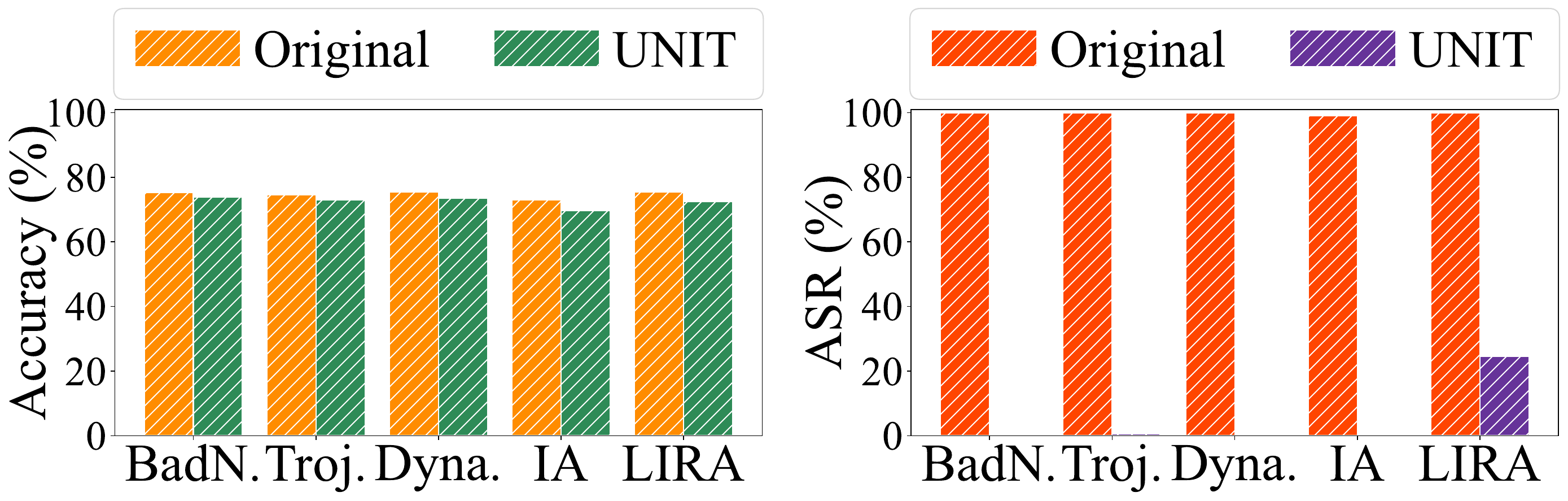}
        \centering (c) CIFAR-100 and Densenet
    \end{minipage}
    \begin{minipage}{0.48\textwidth}
        \includegraphics[width=1\textwidth]{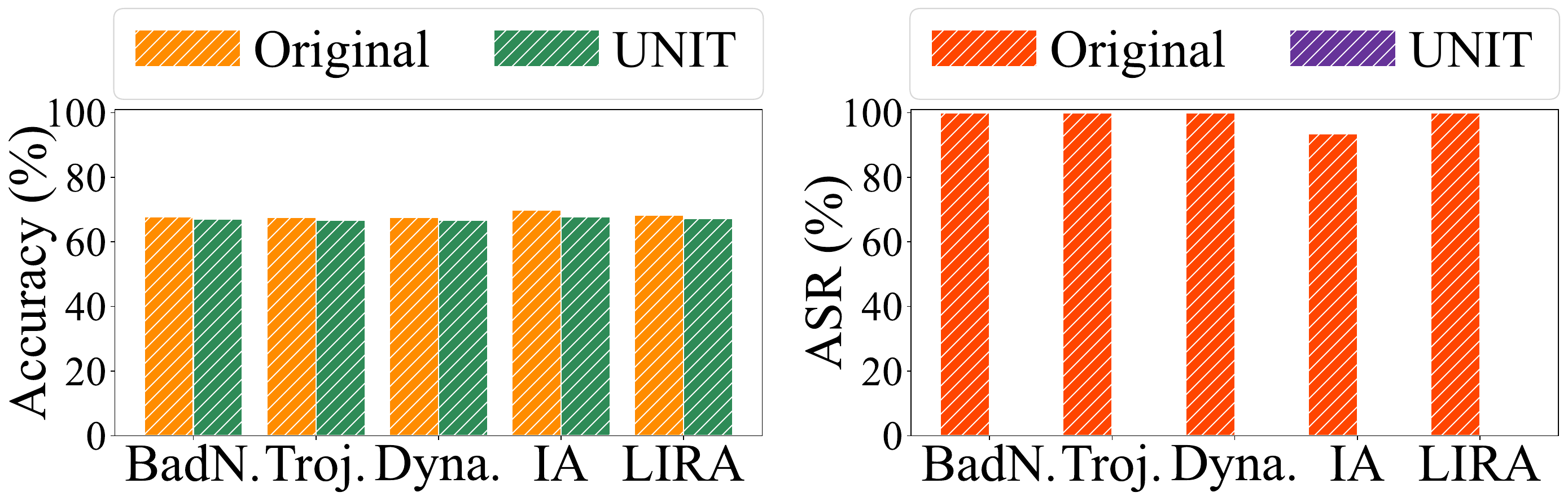}
        \centering (d) CIFAR-100 and Mobilenet
    \end{minipage}
    \begin{minipage}{0.48\textwidth}
        \includegraphics[width=1\textwidth]{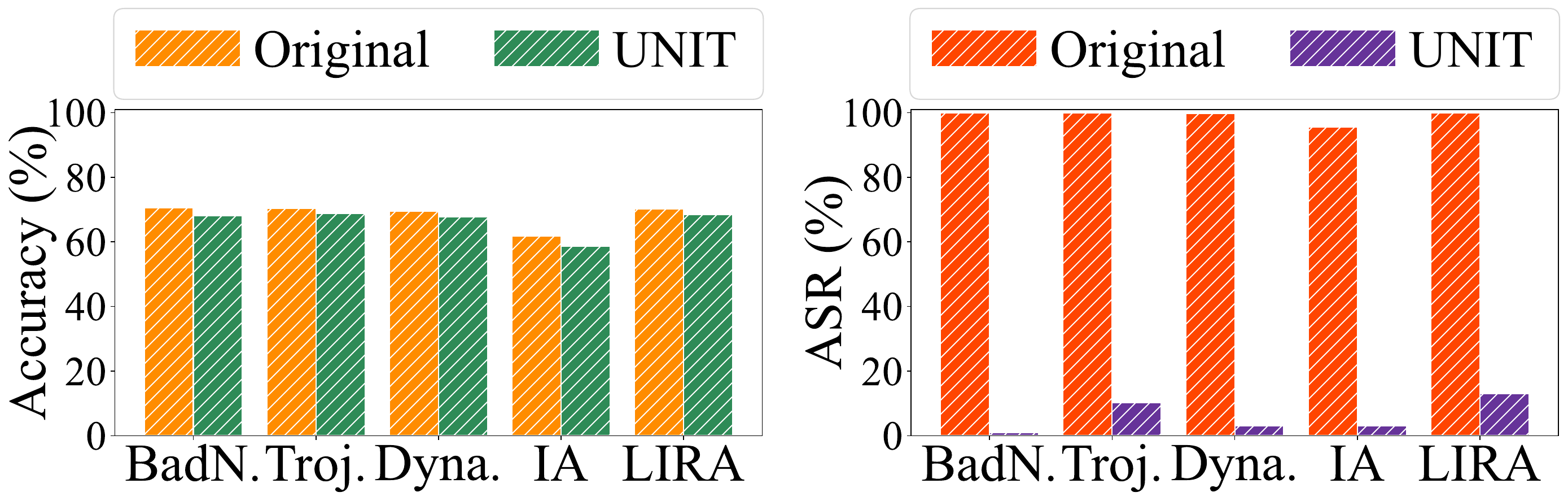}
        \centering (e) STL-10 and WideResNet\
    \end{minipage}
    \begin{minipage}{0.48\textwidth}
        \includegraphics[width=1\textwidth]{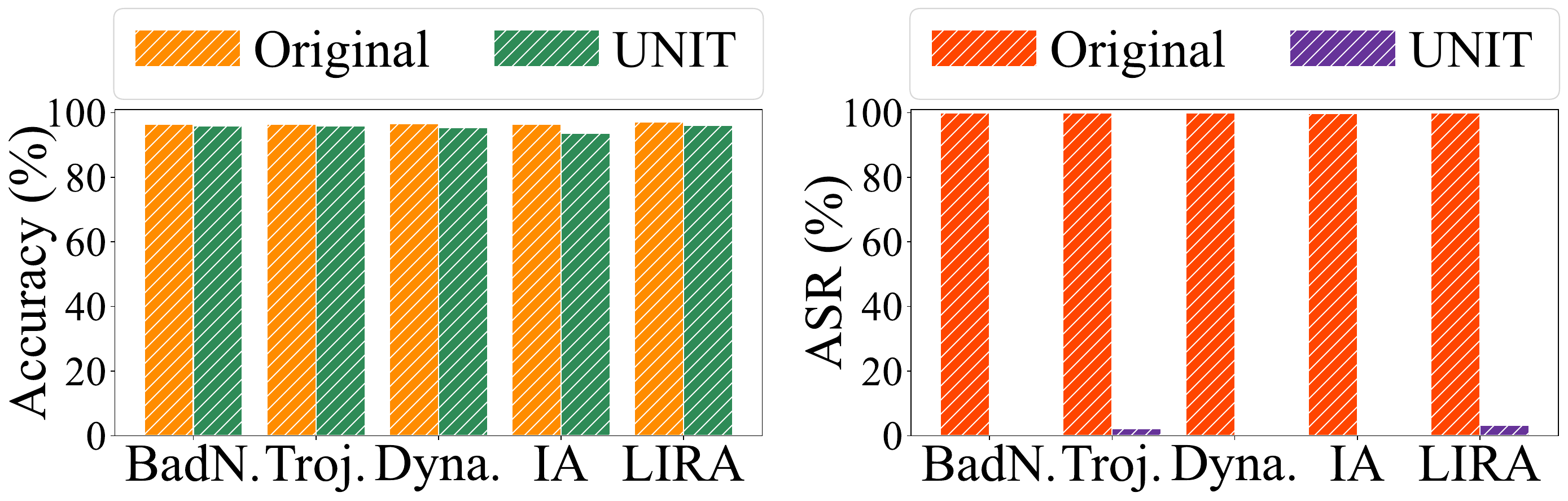}
        \centering (f) GTSRB and PRN34
    \end{minipage}
    \caption{Evaluation on different datasets and network architectures}
    \label{fig:dataset_network}
\end{figure*}

\subsubsection{Evaluation on Various Datasets and Networks}
We extend the evaluation of \Tech{} to include a diverse set of datasets and network architectures. The experiments include four datasets: CIFAR-10~\cite{cifar10}, CIFAR-100~\cite{cifar10}, STL-10~\cite{stl10}, and GTSRB~\cite{gtsrb}, and six network architectures: VGG11~\cite{vgg}, ResNet18~\cite{resnet}, Densenet~\cite{densenet}, Mobilenet~\cite{mobilenet}, WideResNet~\cite{wrn}, and Pre-activation ResNet34 (PRN34)~\cite{resnet}. 5\% of the clean training data is used for defense.
Results are presented in Figure~\ref{fig:dataset_network}, with each sub-figure depicting the outcomes for a specific dataset-network pair. In each sub-figure, the left plot illustrates clean accuracy, while the right plot displays the ASR. The x-axis represents different backdoor attacks, and the y-axis denotes accuracy or ASR. Bar colors in the legend distinguish results before and after the defense.
Notably, \Tech{} consistently reduces ASR from 100\% to near 0\% across various datasets and network architectures. Clean accuracy degradation is minimal in most cases, demonstrating the general effectiveness of \Tech{} across diverse scenarios.

\noindent \underline{\textbf{Application on Transformers.}}
Although \Tech{} is primarily designed for CNN models, we investigate its performance in eliminating backdoor effects in transformers. We poison the CIFAR-10 dataset with BadNets~\cite{badnet} triggers and finetune the ViT-base-patch16-224~\cite{vit} model on it. The model achieves 98.44\% accuracy and 100\% ASR. We then apply \Tech{} to tighten the benign distribution boundary on each \textit{attention layer}, which successfully reduces the ASR to 5.78\%, with a slight accuracy drop of 3.23\%. These results highlight Tech's potential utility in protecting transformers from backdoor attacks.

\subsection{Defense Efficiency} \label{sec:efficient}
We conducted a study on the time cost of various defenses, and the results are illustrated in Figure~\ref{fig:time_cost}. The x-axis represents different methods, and the y-axis indicates the time cost measured in seconds, with each bar denoting the average time cost.
Notably, \Tech{} completes its process in approximately 20 seconds as it only needs to estimate the benign activation distributions based on a small set of clean samples.
Other cost-efficient methods, such as I-BAU and fine-tuning, exhibit similar time costs to \Tech{}. However, as shown in Table~\ref{tab:baseline}, they fall short in defending against a few advanced attacks.

Since \Tech{} modifies the activation layers, we also measure its impact on the model inference. We feed the whole test set containing 10,000 images to the ResNet18 model on CIFAR-10 before and after applying \Tech{}.
The experiment is repeated 5 times.
The time cost is $2.79 \pm 0.35$s for the original model, and $2.86 \pm 0.20$s for the model integrated with \Tech{}.
The inference time difference is negligible (around 2.5\%).
Such a small increase during inference is acceptable as \Tech{} can effectively preclude all evaluated backdoor attacks.

\subsection{Impact on Clean Models} \label{sec:clean_effect}
\begin{figure}[t]
    \centering
    \begin{minipage}[c]{0.5\textwidth}
        \centering
        \includegraphics[width=1\textwidth]{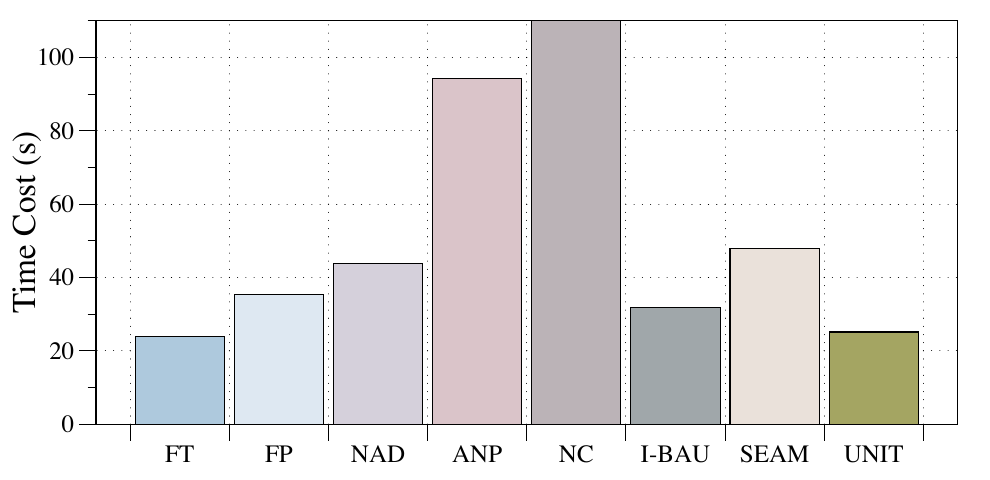}
        \caption{Time cost of different baselines}
        \label{fig:time_cost}
    \end{minipage}
    \hfill
    \begin{minipage}[c]{0.47\textwidth}
        \centering
        \tiny
        \tabcolsep=2.7pt
        \captionof{table}{Impact on clean models}
        \label{tab:clean_models}
        \begin{tabular}{llccc}
             \toprule
             \textbf{Dataset} & \textbf{Network} & \textbf{Original} & \textbf{\Tech{}} & \textbf{Diff.} \\
             \midrule
             CIFAR-10 & VGG11 & 91.88\% & 89.93\% & 1.95\% \\
             CIFAR-10 & ResNet18 & 95.08\% & 92.25\% & 2.83\% \\
             CIFAR-100 & Densenet & 74.78\% & 73.07\% & 1.71\% \\
             CIFAR-100 & Mobilenet & 67.74\% & 67.42\% & 0.32\% \\
             STL-10 & WideResNet & 69.81\% & 69.45\% & 0.36\% \\
             GTSRB & PRN34 & 97.12\% & 95.21\% & 1.91\% \\
             \bottomrule
        \end{tabular}
    \end{minipage}
\end{figure}

We investigate the impact of \Tech{} on clean models, considering that defenders may apply \Tech{} without prior knowledge of whether a model is poisoned. Table~\ref{tab:clean_models} presents the accuracy before and after applying \Tech{} on various clean models. Notably, the degradation of clean accuracy ranges from 0.32\% to 2.83\%, highlighting the minimal impact.



\subsection{Additional Evaluation of \Tech{}} \label{sec:additional_exp}
We conduct evaluation on the latest backdoor attacks and compare \Tech{} with a few more recent defenses in Appendix~\ref{sec:appendix_baseline}.
We evaluate \Tech{}'s performance under three adaptive attack scenarios in Appendix~\ref{sec:appendix_adaptive}, showing its robustness against them.
We carry out a series of ablation studies to examine \Tech{}'s resilience across various hyper-parameters and attack settings in Appendix~\ref{sec:appendix_ablation}. 

\section{Conclusion} \label{sec:conclusion}
We present a novel backdoor mitigation technique designed to approximate a tight distribution for each neuron.
It then effectively reduce any high activation that exceeds the established boundary.
Our comprehensive evaluation illustrates the high efficacy of \Tech{}, outperforming 7 baselines across 14 existing attacks.

\section*{Acknowledgements}
We thank the anonymous reviewers for their constructive comments. We are grateful to the Center for AI Safety for providing computational resources. This research was supported, in part by IARPA TrojAI W911NF-19-S0012, NSF 1901242 and 1910300, ONR N000141712045, N000141410468 and N000141712947. Any opinions, findings, and conclusions in this paper are those of the authors only and do not necessarily reflect the views of our sponsors.

\bibliographystyle{splncs04}
\bibliography{reference}

\clearpage
\newpage

\appendix
\section{Appendix}

\subsection{Details of Baseline Backdoor Attacks and Defenses} \label{sec:appendix_details}
In this section, we provide the details of leveraged baseline backdoor attacks and defense methods.

\begin{figure}[ht]
    \centering
    \footnotesize
    \begin{minipage}{0.13\textwidth}
        \includegraphics[width=1.0\textwidth]{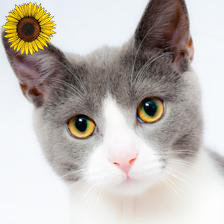}
    \end{minipage}
    \begin{minipage}{0.13\textwidth}
        \includegraphics[width=1.0\textwidth]{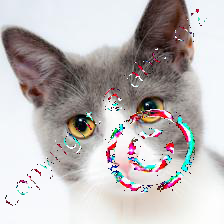}
    \end{minipage}
    \begin{minipage}{0.13\textwidth}
        \includegraphics[width=1.0\textwidth]{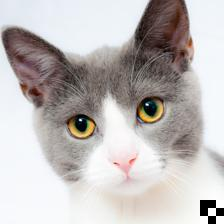}
    \end{minipage}
    \begin{minipage}{0.13\textwidth}
        \includegraphics[width=1.0\textwidth]{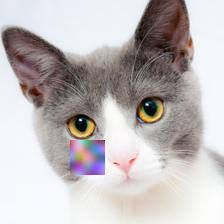}
    \end{minipage}
    \begin{minipage}{0.13\textwidth}
        \includegraphics[width=1.0\textwidth]{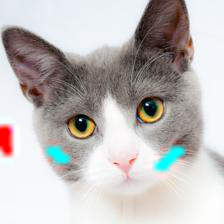}
    \end{minipage}
    \begin{minipage}{0.13\textwidth}
        \includegraphics[width=1.0\textwidth]{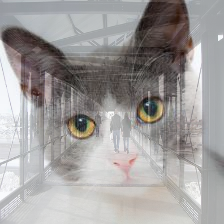}
    \end{minipage}
    \begin{minipage}{0.13\textwidth}
        \includegraphics[width=1.0\textwidth]{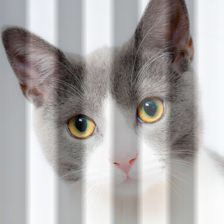}
    \end{minipage}
    \\[\baselineskip]
    \begin{minipage}{0.13\textwidth}
        \includegraphics[width=1.0\textwidth]{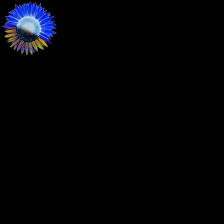}
        \centering \scriptsize BadNets
    \end{minipage}
    \begin{minipage}{0.13\textwidth}
        \includegraphics[width=1.0\textwidth]{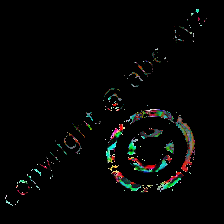}
        \centering \scriptsize Trojan
    \end{minipage}
    \begin{minipage}{0.13\textwidth}
        \includegraphics[width=1.0\textwidth]{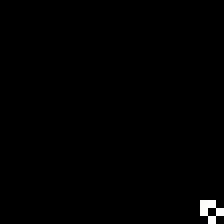}
        \centering \scriptsize CL
    \end{minipage}
    \begin{minipage}{0.13\textwidth}
        \includegraphics[width=1.0\textwidth]{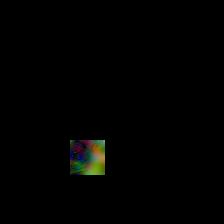}
        \centering \scriptsize Dynamic
    \end{minipage}
    \begin{minipage}{0.13\textwidth}
        \includegraphics[width=1.0\textwidth]{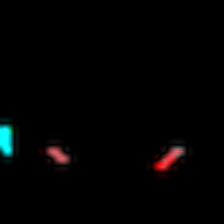}
        \centering \scriptsize IA
    \end{minipage}
    \begin{minipage}{0.13\textwidth}
        \includegraphics[width=1.0\textwidth]{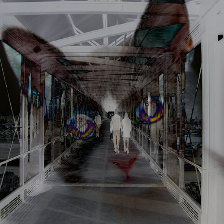}
        \centering \scriptsize Reflection
    \end{minipage}
    \begin{minipage}{0.13\textwidth}
        \includegraphics[width=1.0\textwidth]{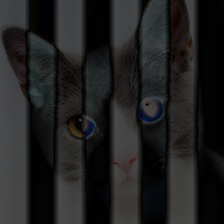}
        \centering \scriptsize SIG
    \end{minipage}
    \\[\baselineskip]
    \vspace{3pt}
    \begin{minipage}{0.13\textwidth}
        \includegraphics[width=1.0\textwidth]{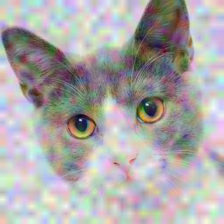}
    \end{minipage}
    \begin{minipage}{0.13\textwidth}
        \includegraphics[width=1.0\textwidth]{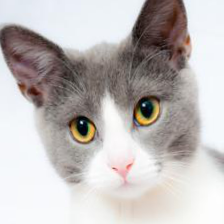}
    \end{minipage}
    \begin{minipage}{0.13\textwidth}
        \includegraphics[width=1.0\textwidth]{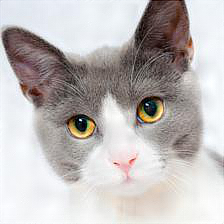}
    \end{minipage}
    \begin{minipage}{0.13\textwidth}
        \includegraphics[width=1.0\textwidth]{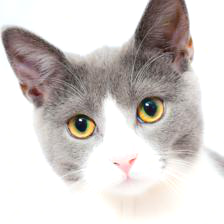}
    \end{minipage}
    \begin{minipage}{0.13\textwidth}
        \includegraphics[width=1.0\textwidth]{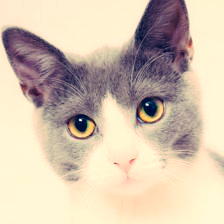}
    \end{minipage}
    \begin{minipage}{0.13\textwidth}
        \includegraphics[width=1.0\textwidth]{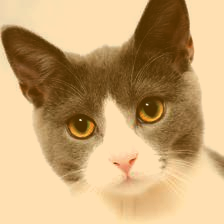}
    \end{minipage}
    \begin{minipage}{0.13\textwidth}
        \includegraphics[width=1.0\textwidth]{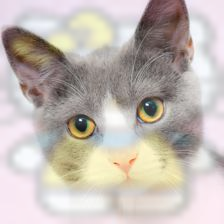}
    \end{minipage}
    \\[\baselineskip]
    \begin{minipage}{0.13\textwidth}
        \includegraphics[width=1.0\textwidth]{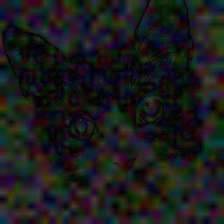}
        \centering \scriptsize Blend
    \end{minipage}
    \begin{minipage}{0.13\textwidth}
        \includegraphics[width=1.0\textwidth]{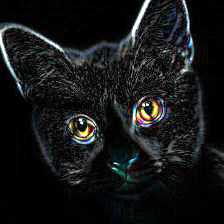}
        \centering \scriptsize WaNet
    \end{minipage}
    \begin{minipage}{0.13\textwidth}
        \includegraphics[width=1.0\textwidth]{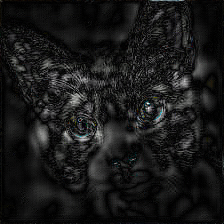}
        \centering \scriptsize ISSBA
    \end{minipage}
    \begin{minipage}{0.13\textwidth}
        \includegraphics[width=1.0\textwidth]{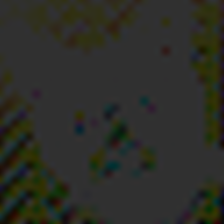}
        \centering \scriptsize LIRA
    \end{minipage}
    \begin{minipage}{0.13\textwidth}
        \includegraphics[width=1.0\textwidth]{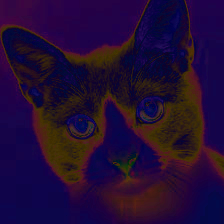}
        \centering \scriptsize Instagram
    \end{minipage}
    \begin{minipage}{0.13\textwidth}
        \includegraphics[width=1.0\textwidth]{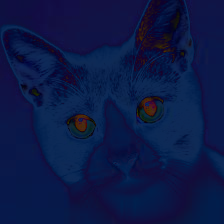}
        \centering \scriptsize DFST
    \end{minipage}
    \begin{minipage}{0.13\textwidth}
        \includegraphics[width=1.0\textwidth]{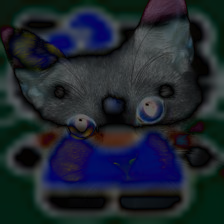}
        \centering \scriptsize Adap-Blend
    \end{minipage}
    \caption{Backdoor trigger examples}
    \label{fig:intro_trigger}
\end{figure}

\smallskip \noindent
\textbf{Backdoor Attacks.}
We evaluate on 14 prominent backdoor attacks, following the original trigger patterns and poisoning strategies, with a fixed poisoning rate of 10\%. Figure~\ref{fig:intro_trigger} provides visualizations of different backdoor triggers, where we displays images stamped with triggers in the first rows and the differences between poisoned images and their source versions in the second rows.

\begin{itemize}
    \item \textit{BadNets}~\cite{badnet} introduces backdoor attacks by incorporating a small percentage of poisoned samples into the training data using standard data-poisoning. During inference, images stamped with the trigger are misclassified to the specified target label.
    \item \textit{Trojan}~\cite{trojannn} injects the backdoor by manipulating selective internal neurons, ensuring the trigger activates these neurons with high values, causing targeted misclassification.
    \item \textit{CL}~\cite{clean_label} proposes a clean-label attack that poisons only samples of the target class during training. It introduces adversarial perturbations to target inputs, misclassifying them to other classes before applying data-poisoning.
    \item \textit{Dynamic backdoor}~\cite{dynamic} leverages a trigger generator to inject various triggers randomly into different inputs.
    \item \textit{IA}~\cite{inputaware} utilizes two trigger generation networks to create trigger masks and patterns based on the inputs, establishing a unique one-to-one mapping between the input and its trigger.
    \item \textit{Reflection}~\cite{reflection} blends inputs with another image to produce a reflection effect.
    \item \textit{SIG}~\cite{sig} perturbs input images with strip effects. It also operates as a clean-label attack.
    \item \textit{Blend}~\cite{blend} applies small random perturbations and blends them with the input to create the trigger.
    \item \textit{WaNet}~\cite{wanet} uses a complex wrapping function to induce a line-bending effect on inputs as the trigger.
    \item \textit{ISSBA}~\cite{invisible} injects an invisible trigger using an image-to-image transforming network.
    \item \textit{LIRA}~\cite{lira} employs a network to inject sample-specific perturbations into inputs as the trigger.
    \item \textit{Instagram}~\cite{abs} uses Instagram filters to introduce the trigger.
    \item \textit{DFST}~\cite{dfst} utilizes a CycleGAN to apply a sunshine effect on inputs as the trigger. It also incorporates a detoxification process to eliminate low-level trigger features, directing the model's focus to high-level features.
    \item \textit{Adap-Blend}~\cite{adaptive_blend} leverages asymmetric and low-confidence training to reduce the latent distance between clean and poisoned samples, enhancing the stealthiness and robustness of the attack against existing defenses.
\end{itemize}

\smallskip \noindent
\textbf{Backdoor Mitigation Baselines.}
We compare our technique \Tech{} with 7 state-of-the-art defenses given the same number of clean training data. We follow the original implementation to conduct experiments and tuning parameters to acquire best performance.
\begin{itemize}
    \item \textit{Standard Fine-Tuning (FT)} retrains the model using the given clean data. We perform fine-tuning for 20 epochs with an initial learning rate of $10^{-2}$ and reduce the learning rate by a factor of 10 every 4 epochs. Data augmentation techniques, including random cropping, rotation, and horizontal flipping, are applied to enhance model generalization.
    \item \textit{Fine-Pruning (FP)}~\cite{fine_pruning} first prunes dormant neurons with low activation values on clean inputs (potential backdoor neurons) and then applies standard fine-tuning to the model.
    \item \textit{NAD}~\cite{nad} distills the knowledge from the teacher model to the student model. The teacher model is derived from the backdoored model after standard fine-tuning. The backdoor effect is removed through the distillation to the student model, only based on clean representations.
    \item \textit{ANP}~\cite{anp} observes that backdoor neurons are sensitive to small perturbations in weight values. It prunes the most sensitive neurons based on this observation.
    \item \textit{NC}~\cite{nc} reverse-engineers backdoor triggers and applies adversarial training to neutralize the effectiveness of the generated triggers.
    \item \textit{I-BAU}~\cite{i-bau} introduces a min-max formulation to eliminate backdoors and leverages implicit hypergradients to optimize the balance between removal efficiency and effectiveness.
    \item \textit{SEAM}~\cite{seam} leverages the catastrophe forgetting assumption~\cite{catastrophe} by first retraining the model on clean samples with randomly assigned labels to forget both clean and backdoor behaviors. It then fine-tune the model on samples with the correct labels to restore the clean performance.
\end{itemize}

\begin{table}[h]
    \centering
    \tiny
    \tabcolsep=2.5pt
    \caption{Evaluation results on the latest backdoor attacks and defenses}
    \label{tab:appendix_more_baselines}
    \begin{tabular}{lgrgrgrgrgrgrgr}
         \toprule
         \multirow{2}{*}{\textbf{Attacks}} & \multicolumn{2}{c}{\textbf{No Defense}} & \multicolumn{2}{c}{\textbf{CLP}} & \multicolumn{2}{c}{\textbf{FST}} & \multicolumn{2}{c}{\textbf{RNP}} & \multicolumn{2}{c}{\textbf{FT-SAM}} & \multicolumn{2}{c}{\textbf{Super-FT}} & \multicolumn{2}{c}{\textbf{\Tech{}}} \\
         \cmidrule(lr){2-3} \cmidrule(lr){4-5} \cmidrule(lr){6-7} \cmidrule(lr){8-9} \cmidrule(lr){10-11} \cmidrule(lr){12-13} \cmidrule(lr){14-15}
         ~ & \cellcolor{white}{Acc.} & \cellcolor{white}{ASR} & \cellcolor{white}{Acc.} & \cellcolor{white}{ASR} & \cellcolor{white}{Acc.} & \cellcolor{white}{ASR} & \cellcolor{white}{Acc.} & \cellcolor{white}{ASR} & \cellcolor{white}{Acc.} & \cellcolor{white}{ASR} & \cellcolor{white}{Acc.} & \cellcolor{white}{ASR} & \cellcolor{white}{Acc.} & \cellcolor{white}{ASR} \\
         \midrule
         BadNets    & 94.82 & 100.0 & 92.37 & 1.18 & 92.50 & \textbf{0.00} & 92.18 & 0.64 & 91.75 & 0.86 & 91.58 & 0.90 & \textbf{92.48} & 0.78 \\
         Instagram  & 94.62 & 99.59 & 91.69 & 48.08 & \textbf{91.93} & 8.55 & 90.05 & 14.50 & 90.96 & 9.91 & 90.55 & 6.83 & 91.43 & \textbf{4.98} \\
         Reflection & 93.29 & 99.59 & \textbf{92.06} & 36.16 & 91.46 & 7.58 & 89.59 & 84.30 & 90.58 & 53.96 & 91.02 & 48.07 & 91.44 & \textbf{6.63} \\
         \midrule
         NARCISSUS  & 92.67 & 95.48 & 87.98 & 84.64 & 86.86 & 41.49 & 89.17 & 79.96 & 89.12 & 43.61 & \textbf{89.31} & 59.48 & 88.74 & \textbf{37.97} \\
         COMBAT     & 94.00 & 88.74 & 90.07 & 50.93 & 91.66 & 73.79 & 90.99 & 72.57 & 89.22 & 62.42 & \textbf{90.61} & 59.31 & 90.13 & \textbf{48.58} \\
         \bottomrule
    \end{tabular}
\end{table}

\subsection{Additional Evaluation on the Latest Backdoor Attacks and Defense Mechanisms} \label{sec:appendix_baseline}
We compare \Tech{} with five additional state-of-the-art baselines, CLP~\cite{clp}, FST~\cite{fst}, RNP~\cite{rnp}, FT-SAM~\cite{ft-sam} and Super-FT~\cite{super-ft}.
CLP identifies and prunes backdoor neurons by examining the channel Lipschitzness to reduce the backdoor effect. It is based on the observation that backdoor neurons tend to have high channel Lipschitz values.
FST actively deviates the weights of the classification layer (e.g., the last fully connected layer in ResNet-18) from the originally compromised weights. It then fine-tunes the feature extraction weights to calibrates the shifted classification weights, aiming to destroy the backdoor correlation.
RNP identifies malicious neurons by unlearning using clean samples with randomly shuffled labels and then recovering using the ground-truth labels. Malicious neurons stand out as they are sensitive to this change and RNP prunes them accordingly.
FT-SAM leverages sharpness-aware minimization to achieve better unlearning while Super-FT design a special learning rate scheduler to enhance the backdoor unlearning.
We consider three typical backdoor attacks, BadNets~\cite{badnet}, Instagram filter~\cite{abs} and Reflection~\cite{reflection}.
In addition, we include two latest backdoor attacks, NARCISSUS~\cite{narcissus} and COMBAT~\cite{combat}, which are both robust clean-label attacks.
The experiments are conducted using the ResNet-18 model and CIFAR-10 dataset, with 5\% of the original training samples available for defense.
Results, presented in Table~\ref{tab:appendix_more_baselines}, indicate that while all defense techniques are effective against conventional attacks like BadNets, they perform less effectively against more complex and recent attacks, particularly NARCISSUS and COMBAT. \Tech{} consistently surpasses other state-of-the-art methods in mitigating backdoor effects, at the comparable cost to clean accuracy.
However, it is important to note that while \Tech{} demonstrates superior performance, it still falls short against the latest clean-label attacks. This may be due to the extremely subtle distribution differences between poisoned and clean activations introduced by these attacks, making them difficult for \Tech{} to approximate.
Our future work will focus on improving \Tech{}'s effectiveness against these robust clean-label attacks.

\begin{figure}[ht]
    \centering
    \begin{minipage}[c]{0.48\textwidth}
        \centering
        \includegraphics[width=1\textwidth]{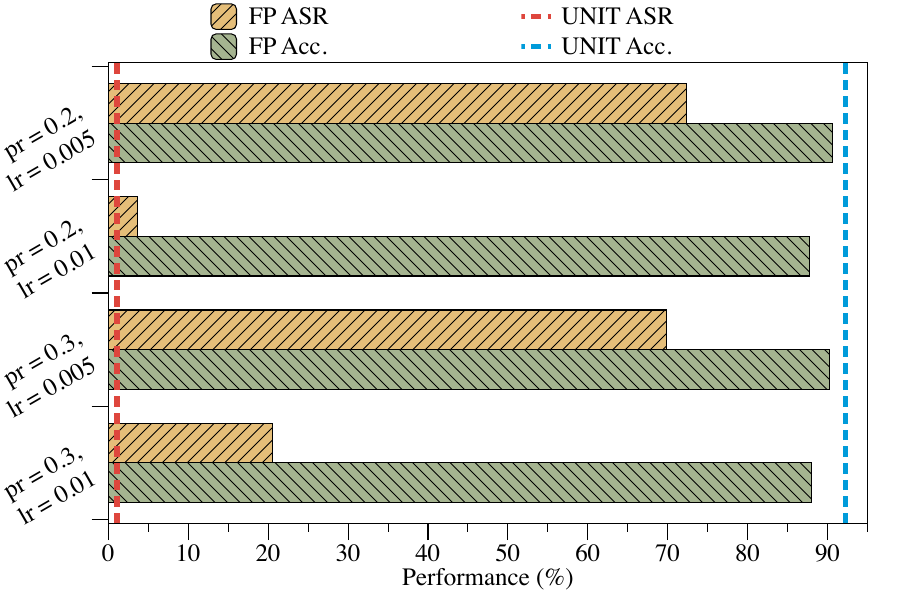}
        \caption{Parameter sensitivity of FP}
        \label{fig:param_fp}
    \end{minipage}
    \hfill
    \begin{minipage}[c]{0.48\textwidth}
        \centering
        \includegraphics[width=1\textwidth]{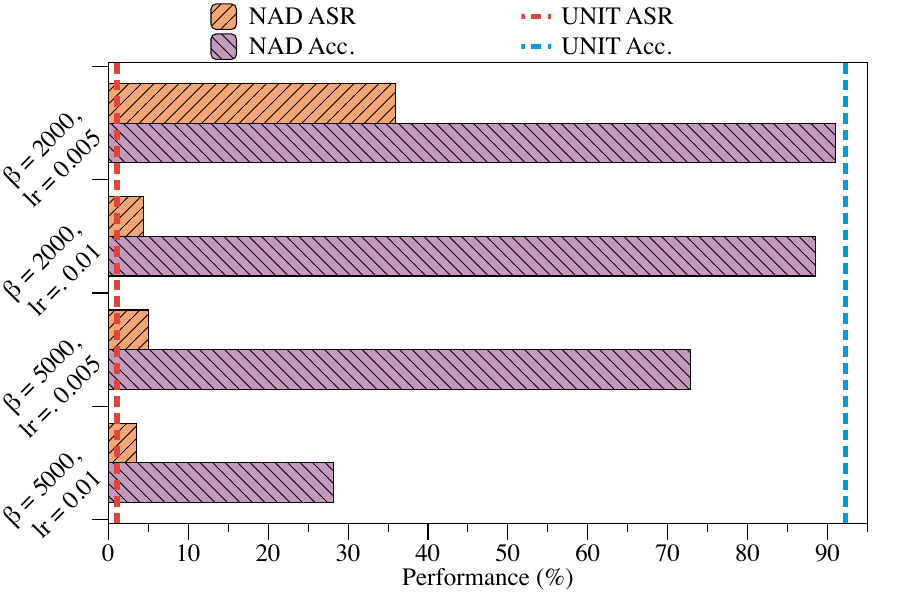}
        \caption{Parameter sensitivity of NAD}
        \label{fig:param_nad}
    \end{minipage}
    \hfill
    \begin{minipage}[c]{0.48\textwidth}
        \centering
        \includegraphics[width=1\textwidth]{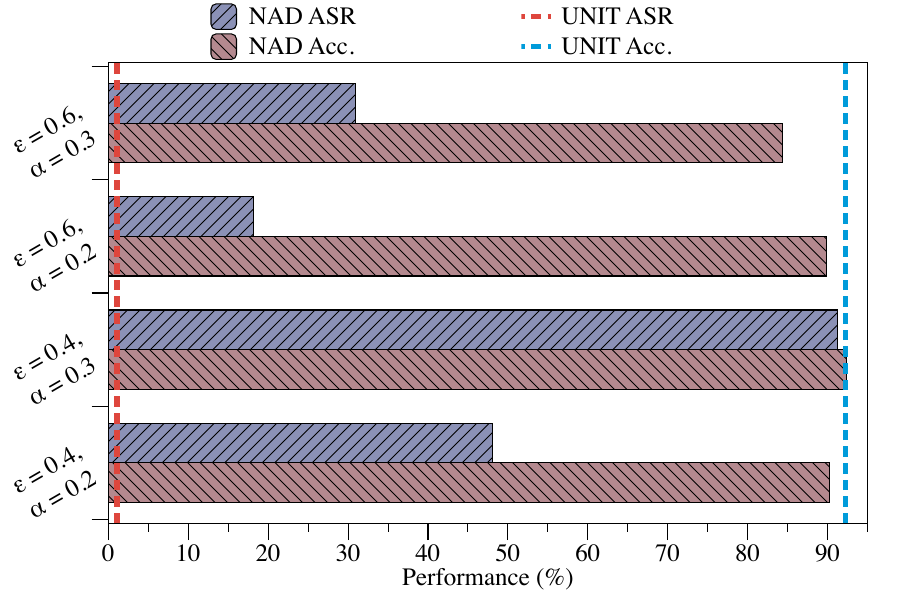}
        \caption{Parameter sensitivity of ANP}
        \label{fig:param_anp}
    \end{minipage}
    \hfill
    \begin{minipage}[c]{0.48\textwidth}
        \centering
        \includegraphics[width=0.94\textwidth]{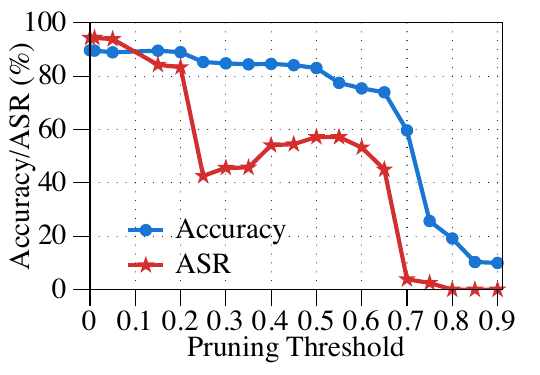}
        \caption{Different pruning ratios}
        \label{fig:rnp_pruning}
    \end{minipage}
\end{figure}

\subsection{Parameter Sensitivity Analysis} \label{sec:appendix_param}
We take three typical baselines to illustrate that existing methods are sensitive to their parameters and require sophisticated parameter tuning to ensure the good performance. However, \Tech{} is parameter-efficient and outperforms the existing methods.
We conduct experiments using CIFAR-10 dataset and ResNet-18 network. We inject CL~\cite{clean_label} backdoor into the model and apply FP~\cite{fine_pruning}, NAD~\cite{nad} and ANP~\cite{anp} to mitigate the attack.
For each defense, we take two key parameters and evaluate the performance for different parameter values.
Specifically, we take the pruning ratio (\textit{pr}) and learning rate (\textit{lr}) for FP, distillation strength ($\beta$) and learning rate (\textit{lr}) for NAD, and adversarial perturbation ($\epsilon$) and pruning coefficient ($\alpha$) for ANP.
Results are presented in Figure~\ref{fig:param_fp}, Figure~\ref{fig:param_nad} and Figure~\ref{fig:param_anp}, where the y-axis denotes the parameter values while the x-axis presents the performance in percentage (accuracy or ASR).
For each parameter setting, we visualize the performance using two bars, i.e., resulting ASR (top bar) and Acc. (lower bar).
The red and blue dashed lines represents the resulting ASR and accuracy after applying \Tech{}.
Observe that the performance of existing methods are significantly sensitive to parameter tuning, with large fluctuation over slightly different parameters.
On the contrary, \Tech{} is parameter-efficient and outperforms the three baselines, indicating its practical applicability.

In addition, we take the state-of-the-art pruning method RNP~\cite{rnp} to defend against the Reflection backdoor~\cite{reflection}.
Figure~\ref{fig:rnp_pruning} shows that as the pruning threshold increases (more neurons are pruned), both accuracy (blue curve) and ASR (red curve) degrade similarly.
This indicates that some neurons handle both clean and backdoor tasks, and no matter how the parameters are tuned to control the pruning rate, it is fundamentally difficult to eliminate the backdoor effect without a non-trivial accuracy cost (Section~\ref{sec:motivation}).
In contrast, \Tech{} achieves 91.44\% accuracy and reduces ASR to 6.63\% by precisely tightening the neural distribution.


\subsection{Adaptive Attacks} \label{sec:appendix_adaptive}
In this section, we discuss three adaptive attack scenarios in detail.

\smallskip \noindent
\textbf{Activation Suppression.}
To tamper \Tech{}'s effectiveness, an adversary may attempt to bridge the gap between benign and backdoor activation. Specifically, an adaptive loss is incorporated during training to suppress the backdoor activation:
\begin{equation}
    Loss = \mathcal{L}(M(x), y) + \mathcal{L}(M(x \oplus T), y_T) + \alpha \cdot \sum_{l=1}^{L} ||F^{l}(x) - F^{l}(x \oplus T)||_{2}^{2},
\end{equation}
\noindent Here, $M$ denotes the model, $x$ and $y$ denote clean images and their labels, $x \oplus T$ denote their poisoned versions and $y_T$ is the attack target class.
$F^{l}(x)$ represents the activation value of $x$ at the $l$-th layer as defined in Section~\ref{sec:notation}.
The adaptive loss term $||F^{l}(x) - F^{l}(x \oplus T)||_2^2$ uses Mean Squared Error (MSE) to reduce the difference between benign and poisoned activation. The parameter $\alpha$ controls the trade-off between the adaptive loss term and the normal cross-entropy loss $\mathcal{L}$.
The experiment is conducted on CIFAR-10 and ResNet18 using BadNets as the backdoor attack. We evaluate five $\alpha$ values: 0, 0.1, 1, 10, 100. Table~\ref{tab:adaptive_loss} presents the results, where the first column shows $\alpha$ values, the second and third columns present accuracy and ASR without defense, the fourth column shows the adaptive loss, and the last two columns present accuracy and ASR after applying \Tech{}.
Observe that with the increase of $\alpha$, the adaptive loss decreases, signifying successful activation suppression. However, \Tech{} continues to effectively reduce the ASR to less than 1.07\%.
The reason is that despite the reduced backdoor activation, \Tech{} can still effectively clip the slightly higher malicious activation.

\begin{table}[ht]
    \centering
    \tiny
    \tabcolsep=2pt
    \caption{Evaluation against activation suppression}
    \label{tab:adaptive_loss}
    \begin{tabular}{lgccgcg}
         \toprule
         \multirow{2}{*}{\textbf{$\alpha$}} & \multicolumn{3}{c}{\textbf{No Defense}} & \multicolumn{2}{c}{\textbf{\Tech{}}} \\
         \cmidrule(lr){2-4} \cmidrule(lr){5-6}
         ~ & \cellcolor{white}{Acc.} & \cellcolor{white}{ASR} & \cellcolor{white}{Adap-Loss} & \cellcolor{white}{Acc.} & \cellcolor{white}{ASR} \\
         \midrule
         0   & 94.84\% & 100.00\% & 0.1108 & 93.05\% & 1.07\% \\
         0.1 & 94.25\% & 100.00\% & 0.0433 & 92.52\% & 0.84\% \\
         1   & 94.13\% & 100.00\% & 0.0137 & 92.42\% & 0.73\% \\
         10  & 93.77\% & 100.00\% & 0.0034 & 91.85\% & 1.02\% \\
         100 & 93.42\% & 100.00\% & 0.0015 & 91.35\% & 0.90\% \\
         \bottomrule
    \end{tabular}
\end{table}

\smallskip \noindent
\textbf{Label-specific Backdoor.}
Attackers may employ label-specific strategies to impact the efficacy of \Tech{}. In label-specific backdoor attacks, the backdoor exclusively influences samples belonging to the victim class. Samples stamped with the trigger but not from the victim class will not be misclassified into the attack target class. Consequently, label-specific backdoor attacks rely on normal features and could potentially impact the difference between clean and malicious activation distributions.
We conduct experiments on CIFAR-10 using ResNet18 and utilized BadNets as a representative attack. When poisoning the model, we not only introduce images of the victim class stamped with the trigger and labeled as the target class but also incorporate negative samples to achieve label-specificity. Negative samples consist of images from classes other than the victim class, stamped with the trigger and labeled as their source labels. These negative samples aid the model in learning the correlation between the backdoor trigger and the victim class.
Table~\ref{tab:adaptive_label_spec} presents the results, with the first column denoting the attack victim-target pair, the second and third columns representing clean accuracy and ASR without defense, the fourth column illustrating the effectiveness of label-specificity (ASR of images stamped with the trigger from non-victim classes), and the last two columns displaying the results after applying \Tech{}.
Observe that negative samples facilitate label-specificity, reducing the ASR of non-victim classes from 100\% to nearly 3\%, while the ASR of the victim class remains high at approximately 98\%. After applying \Tech{}, the clean accuracy remains high with only about a 2\% degradation, while the ASR decreases to less than 4.40\%, demonstrating the effectiveness of \Tech{} against label-specific attacks.
The reason is that even if the backdoor relies on benign features, it still necessitates reasonably large activation values to be triggered. \Tech{} identifies and mitigates these large activation values, rendering it effective against label-specific attacks.

\begin{table}[ht]
    \centering
    \tiny
    \tabcolsep=3pt
    \caption{Adaptive attack through leveraging label specificity.}
    \label{tab:adaptive_label_spec}
    \begin{tabular}{cgccgc}
         \toprule
         \multirow{2}{*}{\textbf{V-T Pair}} & \multicolumn{2}{c}{\textbf{No Defense}} & \textbf{Specificity} & \multicolumn{2}{c}{\textbf{\Tech{}}} \\
         \cmidrule(lr){2-3} \cmidrule(lr){4-4} \cmidrule(lr){5-6}
         ~ & \cellcolor{white}{Acc.} & \cellcolor{white}{ASR} & \cellcolor{white}{Non-victim ASR} & \cellcolor{white}{Acc.} & \cellcolor{white}{ASR} \\
         \midrule
         0-9 & 94.62\% & 98.60\% & 100.00\% $\rightarrow$ 2.88\% & 92.31\% & 1.50\% \\
         8-3 & 94.83\% & 98.60\% & 100.00\% $\rightarrow$ 2.51\% & 92.64\% & 0.90\% \\
         2-4 & 94.71\% & 97.90\% & 100.00\% $\rightarrow$ 3.20\% & 92.74\% & 4.40\% \\
         6-5 & 94.89\% & 98.40\% & 100.00\% $\rightarrow$ 3.04\% & 92.60\% & 0.00\% \\
         7-1 & 94.79\% & 98.20\% & 100.00\% $\rightarrow$ 2.06\% & 92.97\% & 0.00\% \\
         \bottomrule
    \end{tabular}
\end{table}

\smallskip \noindent
\textbf{Trigger-specific Backdoor.}
Attackers may exploit trigger-specificity to dynamically impact \Tech{}. Trigger-specificity entails that the backdoor is activated only when a specific pattern is presented in the image. In other words, a ground-truth trigger stamped with some noise will not induce the backdoor effect.
The potential impact on the effectiveness of \Tech{} arises from the model's potential use of benign neurons to extract high-level features of the trigger pattern. This could lead to a reduction in the difference between benign and backdoor activation distributions.
We conducte experiments on CIFAR-10 using ResNet18 and employ BadNets as an example attack, utilizing a yellow flower as the backdoor trigger. In addition to poisoned samples stamped with the trigger and relabeled as the target class, we introduce negative samples to establish trigger-specificity. We added Gaussian noise to the trigger pattern, stamping the noisy pattern onto certain training images while keeping their ground-truth labels. This approach help the model learn the high-level trigger pattern instead of some low-level features.
Table~\ref{tab:adaptive_label_spec} presents the results, where the first column denotes the noise level added to the negative samples, the second and third columns represent clean accuracy and ASR without defense, the fourth column illustrates the effectiveness of trigger-specificity (ASR of images stamped with noisy triggers), and the last two columns display the results after applying \Tech{}.
Observe that the noisy ASR in the fourth column is significantly reduced when the noise level is $0.05, 0.1, 0.5, 1.0$, indicating that negative samples effectively realize trigger-specificity. Notably, in the last two columns, \Tech{} still reduces the ASR from 100\% to nearly 1\%, while maintaining high clean accuracy.
\Tech{} proves effective against trigger-specific attacks because even if the model learns the high-level backdoor trigger pattern, it cannot circumvent the separation between clean and malicious distributions. This discrepancy is leveraged by \Tech{} to mitigate the backdoor effect.

\begin{table}[ht]
    \centering
    \tiny
    \tabcolsep=3pt
    \caption{Adaptive attack through leveraging trigger specificity.}
    \label{tab:adaptive_trigger_spec}
    \begin{tabular}{lgccgc}
         \toprule
         \multirow{2}{*}{\textbf{Level}} & \multicolumn{2}{c}{\textbf{No Defense}} & \textbf{Specificity} & \multicolumn{2}{c}{\textbf{\Tech{}}} \\
         \cmidrule(lr){2-3} \cmidrule(lr){4-4} \cmidrule(lr){5-6}
         ~ & \cellcolor{white}{Acc.} & \cellcolor{white}{ASR} & \cellcolor{white}{Noisy Trigger ASR} & \cellcolor{white}{Acc.} & \cellcolor{white}{ASR} \\
         \midrule
         0.01 & 94.65\% & 100.00\% & 100.00\% $\rightarrow$ 99.98\% & 92.55\% & 1.01\% \\
         0.05 & 94.79\% & 100.00\% & 100.00\% $\rightarrow$ 6.84\% & 92.85\% & 0.86\% \\
         0.1 & 94.70\% & 100.00\% & 100.00\% $\rightarrow$ 0.59\% & 92.58\% & 1.06\% \\
         0.5 & 94.64\% & 100.00\% & 77.08\% $\rightarrow$ 0.64\% & 92.65\% & 0.87\% \\
         1 & 94.99\% & 100.00\% & 36.50\% $\rightarrow$ 0.76\% & 92.65\% & 0.92\% \\
         \bottomrule
    \end{tabular}
\end{table}

\subsection{Ablation Study} \label{sec:appendix_ablation}
In this section, we perform a series of experiments to assess the performance of \Tech{} under different attack and defense settings. Additionally, we conduct an ablation study on the design choices and hyper-parameters of \Tech{}.

\begin{table}[t]
    \centering
    \tiny
    \tabcolsep=5pt
    \caption{Evaluation on different activation functions}
    \label{tab:ablation_diff_acti}
    \begin{tabular}{lgcgc}
         \toprule
         \multirow{2}{*}{\textbf{Activation}} & \multicolumn{2}{c}{\textbf{Original}} & \multicolumn{2}{c}{\textbf{\Tech{}}} \\
         \cmidrule(lr){2-3} \cmidrule(lr){4-5}
         ~ & \cellcolor{white}{Acc.} & ASR & \cellcolor{white}{Accuracy} & ASR \\
         \midrule
         ReLU & 94.82\% & 100.00\% & 92.46\% & 0.78\% \\
         LeakyReLU & 94.17\% & 100.00\% & 92.02\% & 0.96\% \\
         SELU & 90.89\% & 99.93\% & 89.66\% & 1.06\% \\
         ELU & 91.24\% & 99.98\% & 90.52\% & 0.99\% \\
         TanhShrink & 89.80\% & 100.00\% & 89.21\% & 1.13\% \\
         Softplus & 88.08\% & 99.97\% & 87.19\% & 1.58\% \\
         Sigmoid & 80.43\% & 99.80\% & 77.62\% & 5.31\% \\
         Tanh & 90.95\% & 99.98\% & 88.14\% & 6.03\% \\
         \bottomrule
    \end{tabular}
\end{table}

\subsubsection{Different Activation Functions} \label{sec:acti_func}
There are different types of activation functions used in deep neural networks.
We evaluate \Tech{} on BadNets-poisoned models using different activation functions.
CIFAR-10 and ResNet18 are used for the study.
We replace the standard activation function (ReLU) with five commonly used functions, LeakyReLU~\cite{leakyrelu}, SELU~\cite{selu}, ELU~\cite{elu}, TanhShrink~\cite{tanhshrink}, Softplus~\cite{softplus}, Sigmoid~\cite{sigmoid}, and Tanh~\cite{tanh}.
Table~\ref{tab:ablation_diff_acti} shows the results.
\Tech{} is effective in all the cases, reducing the ASR from near 100.00\% to less than 1.6\%.
The impact on the clean accuracy is negligible (less than 2\% degradation).
This is attributed to the distinct separation between clean and poisoned activation distributions, a phenomenon that persists across various activation functions.
Hence, \Tech{} is able to ensure robust performance irrespective of the activation function utilized.

\begin{table*}[t]
    \centering
    \caption{Ablation study on different number given clean training samples}
    \label{tab:ablation_diff_num}
    \tiny
    \tabcolsep=4pt
    \begin{tabular}{lgrgrgrgrgrgr}
        \toprule
        \multirow{2}{*}{Attacks} &\multicolumn{2}{c}{\textbf{No Defense}} &\multicolumn{2}{c}{\textbf{10\%}} &\multicolumn{2}{c}{\textbf{5\%}} &\multicolumn{2}{c}{\textbf{1\%}} &\multicolumn{2}{c}{\textbf{0.1\%}} \\ \cmidrule{2-11}
        & \cellcolor{white}{Acc.} & ASR & \cellcolor{white}{Acc.} & ASR & \cellcolor{white}{Acc.} & ASR & \cellcolor{white}{Acc.} & ASR & \cellcolor{white}{Acc.} & ASR \\
        \cmidrule{1-11}
        BadNets &94.82\% &100.00\% &92.81\% &1.09\% &92.48\% &0.78\% &90.60\% &1.59\% &88.01\% &2.63\% \\
        Trojan &94.73\% &100.00\% &92.44\% &1.65\% &92.38\% &2.17\% &91.07\% &1.96\% &89.93\% &2.73\% \\
        CL &94.58\% &98.46\% &92.36\% &0.93\% &92.21\% &1.09\% &89.98\% &1.81\% &90.90\% &8.09\% \\
        Dynamic &95.08\% &100.00\% &92.34\% &1.28\% &92.77\% &1.54\% &88.48\% &2.01\% &90.29\% &2.14\% \\
        IA &91.15\% &97.96\% &90.78\% &1.82\% &89.93\% &1.03\% &87.66\% &2.11\% &89.38\% &3.45\% \\
        Reflection &93.29\% &99.33\% &91.64\% &2.74\% &91.44\% &6.63\% &88.82\% &7.47\% &83.02\% &0.46\% \\
        SIG &94.97\% &99.80\% &92.45\% &0.41\% &92.48\% &1.74\% &89.30\% &0.46\% &86.88\% &1.18\% \\
        Blend &94.62\% &100.00\% &91.75\% &1.55\% &91.99\% &1.18\% &88.77\% &2.06\% &90.86\% &0.61\% \\
        WaNet &94.36\% &99.80\% &91.09\% &1.83\% &91.02\% &2.44\% &89.53\% &9.90\% &82.96\% &3.68\% \\
        ISSBA &94.55\% &100.00\% &91.96\% &1.19\% &91.84\% &1.57\% &89.53\% &1.86\% &88.10\% &2.53\% \\
        LIRA &95.11\% &100.00\% &92.65\% &0.82\% &92.29\% &0.58\% &89.49\% &1.93\% &82.76\% &1.17\% \\
        Instagram &94.62\% &99.59\% &91.71\% &4.05\% &91.43\% &4.98\% &91.07\% &11.66\% &86.25\% &21.64\% \\
        DFST &93.25\% &99.77\% &91.83\% &3.54\% &91.64\% &4.02\% &88.59\% &3.90\% &88.97\% &8.80\% \\
        Adap-Blend &94.22\% &82.80\% &91.35\% &18.39\% &90.84\% &15.03\% &88.78\% &47.64\% &87.64\% &58.57\% \\
        \midrule
        Average &94.26\% &98.39\% &91.94\% &2.95\% &91.77\% &3.20\% &89.41\% &5.88\% &87.57\% &8.41\% \\
        \bottomrule
    \end{tabular}
\end{table*}

\subsubsection{Different Numbers of Clean Training Data.}
We study the impact of different numbers of given clean training samples on \Tech{}. Table~\ref{tab:ablation_diff_num} presents the results, comparing \Tech{}'s performance when provided with 10\%, 5\% (default), 1\%, and 0.1\% of training data for defense. The experiments are conducted on CIFAR-10 and ResNet-18.
Observe that in general, \Tech{} exhibits better performance (higher clean accuracy and lower ASR) when given more clean samples. This is expected as \Tech{} requires clean samples to approximate a tight distribution and eliminate malicious activation. More samples allowing for a more precise approximation of the real distribution. Obtaining clean samples is generally feasible, even from the internet, supporting the argument that \Tech{} is generally effective and suitable for real-world applications.
Moreover, even with access to only 0.1\% of samples (totally 50 images in CIFAR-10 dataset), \Tech{} still effectively reduces ASR to an average of 8.41\%, albeit with a sacrifice of around 7\% in accuracy.

\begin{table}[ht]
    \centering
    \tiny
    \tabcolsep=5pt
    \caption{Evaluation on different low poisoning rates}
    \label{tab:ablation_low_poison_rate}
    \begin{tabular}{lgrgrgrgrgr}
         \toprule
         \multirow{2}{*}{\textbf{Attacks}} & \multicolumn{2}{c}{\textbf{No Defense}} & \multicolumn{2}{c}{\textbf{FP}} & \multicolumn{2}{c}{\textbf{NAD}} & \multicolumn{2}{c}{\textbf{ANP}} & \multicolumn{2}{c}{\textbf{\Tech{}}} \\
         \cmidrule(lr){2-3} \cmidrule(lr){4-5} \cmidrule(lr){6-7} \cmidrule(lr){8-9} \cmidrule(lr){10-11}
         ~ & \cellcolor{white}{Acc.} & \cellcolor{white}{ASR} & \cellcolor{white}{Acc.} & \cellcolor{white}{ASR} & \cellcolor{white}{Acc.} & \cellcolor{white}{ASR} & \cellcolor{white}{Acc.} & \cellcolor{white}{ASR} & \cellcolor{white}{Acc.} & \cellcolor{white}{ASR} \\
         \midrule
         BadNets (1\%)   & 93.87 & 100.0 & 89.78 & 52.32 & 88.69 & 38.62 & 88.15 & 7.56 & \textbf{92.72} & \textbf{0.85} \\
         BadNets (0.1\%) & 94.05 & 100.0 & 89.66 & 99.84 & 89.72 & 49.05 & 86.45 & 37.27 & \textbf{90.22} & \textbf{1.44} \\
         \midrule
         Blend (1\%)     & 93.92 & 99.92 & 89.43 & 29.69 & 89.64 & 0.07 & 88.16 & 0.10 & \textbf{92.96} & \textbf{0.00} \\
         Blend (0.1\%)   & 93.88 & 99.89 & 88.87 & 41.56 &  88.64 & 0.20 & 87.51 & 0.48 & \textbf{92.14} & \textbf{0.11} \\
         \midrule
         WaNet (1\%)     & 93.59 & 94.91 & 89.01 & 5.77 & \textbf{91.31} & 10.43 & 89.93 & \textbf{1.48} & 90.97 & 2.19 \\
         WaNet (0.1\%)   & 93.80 & 93.75 & 89.79 & 9.89 & 90.22 & 21.39 & \textbf{90.26} & \textbf{1.44} & 90.08 & 4.07 \\
         \bottomrule
    \end{tabular}
\end{table}

\begin{figure}[t]
    \centering
    \begin{minipage}[c]{0.48\textwidth}
        \centering
        \includegraphics[width=1\textwidth]{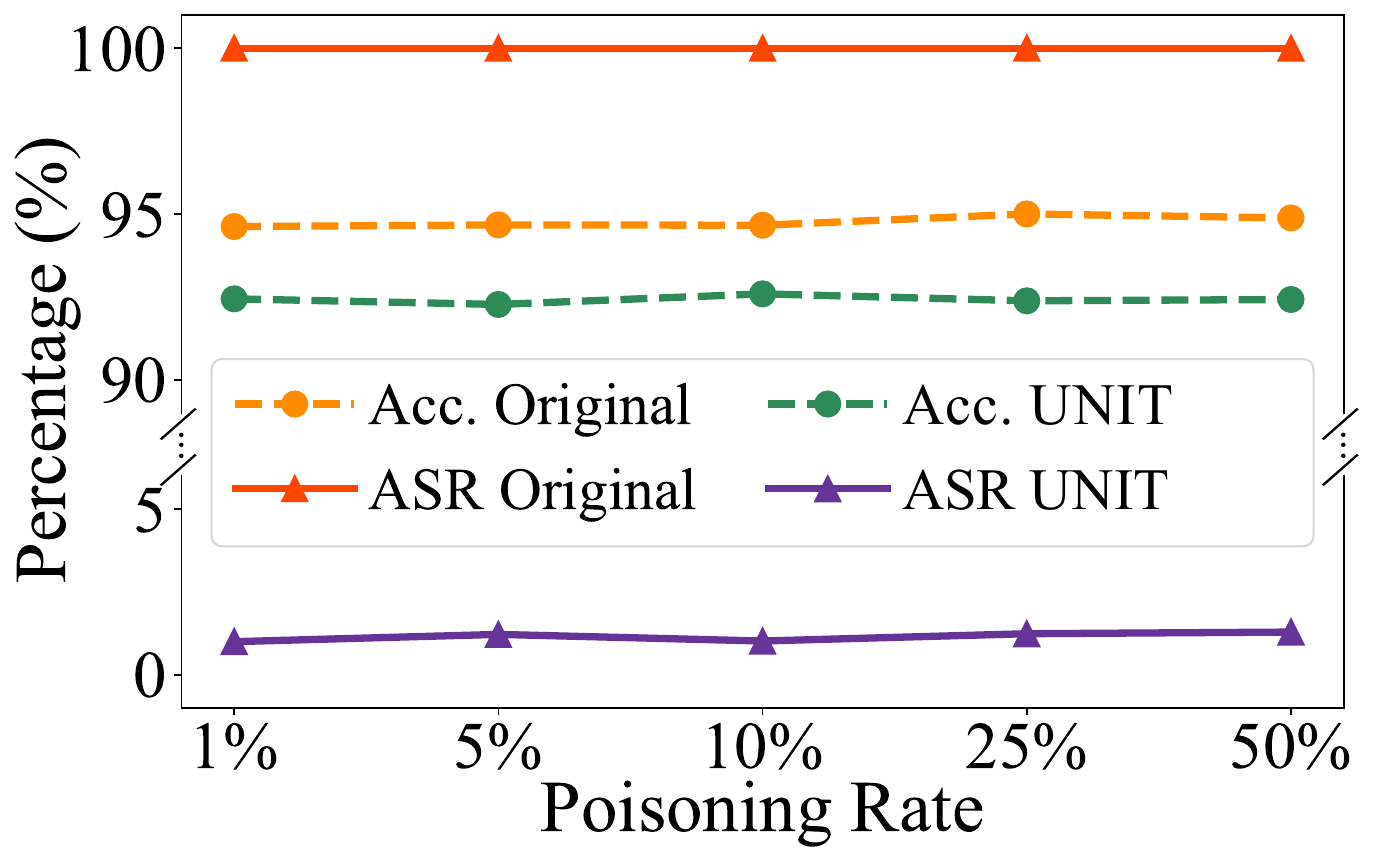}
        \caption{Ablation study on different poisoning rates}
        \label{fig:robust_rate}
    \end{minipage}
    \hfill
    \begin{minipage}[c]{0.48\textwidth}
        \centering
        \includegraphics[width=1\textwidth]{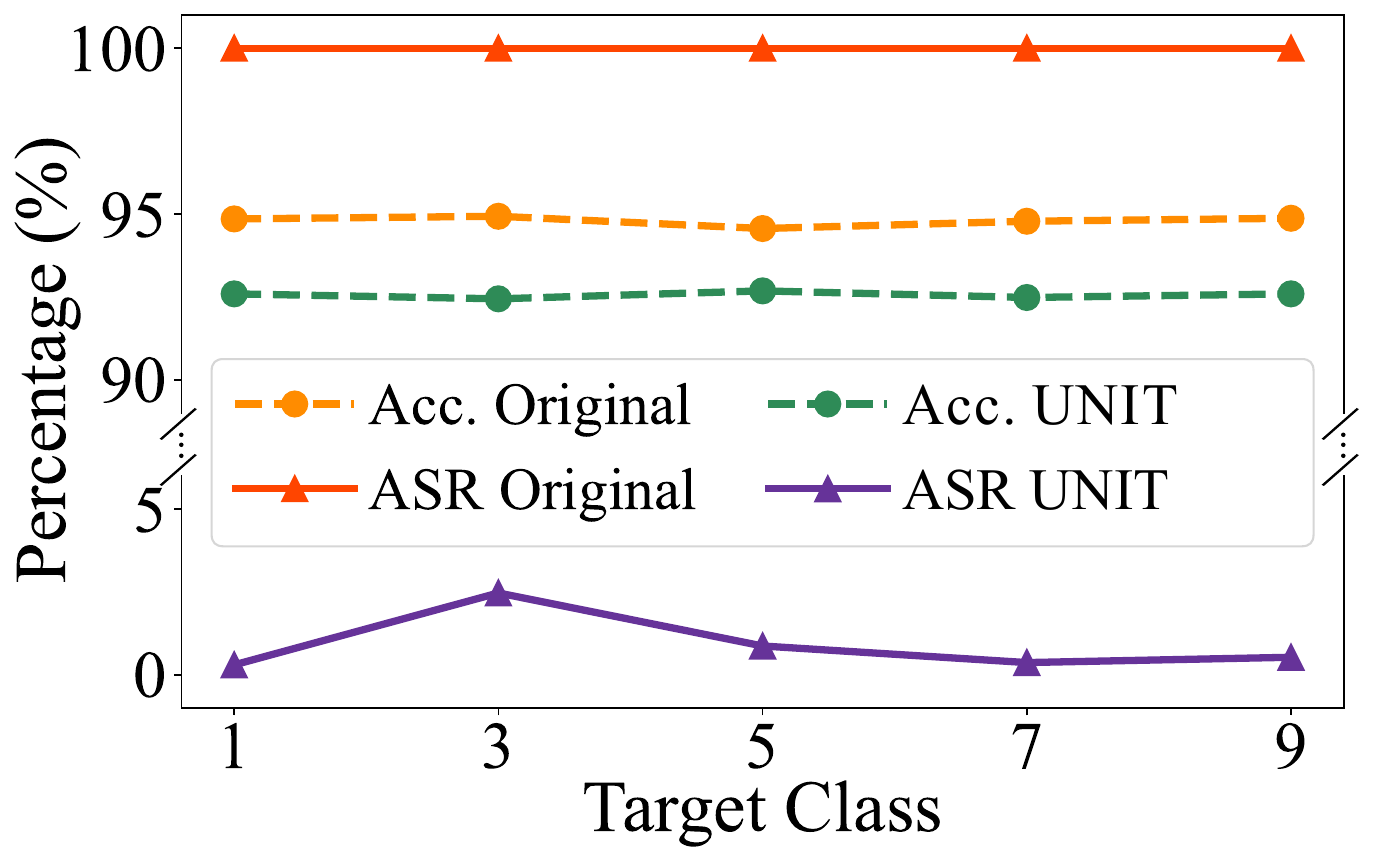}
        \caption{Ablation study on different target labels}
        \label{fig:robust_target}
    \end{minipage}
\end{figure}

\subsubsection{Different Poisoning Rates.}
We study the impact of various poisoning rates on defense performance.
Utilizing CIFAR-10 and ResNet-18, we introduce BadNets triggers with poisoning rates of 1\%, 5\%, 10\%, 25\%, and 50\%. Figure~\ref{fig:robust_rate} illustrates the results, with the x-axis representing poisoning rates and the y-axis denoting performance. Notably, \Tech{} consistently reduces the ASR from 100.00\% to approximately 1\%, with minimal accuracy sacrifice (less than 2\%). This highlights the robustness of \Tech{} across different data poisoning rates.

Additionally, we compared \Tech{} with three baselines on CIFAR-10 and ResNet-18 at extremely low poison rates (1\% and 0.1\%). Table~\ref{tab:ablation_low_poison_rate} shows that while baseline performance degrades at lower poison rates, \Tech{} remains robust and outperforms them.

\subsubsection{Different Target Labels.}
We investigate the influence of different target labels on defense performance of \Tech{}. We use CIFAR-10, ResNet-18 and BadNets trigger to conduct the experiments and employ labels 1, 3, 5, 7, and 9 as the targets. Figure~\ref{fig:robust_target} illustrates the results, where the x-axis represents the target classes, and the y-axis indicates performance.
Remarkably, \Tech{} consistently reduces the ASR from 100.00\% to 0.30\%-2.47\% without significantly impacting clean accuracy, underscoring the robustness of \Tech{} across various target labels.

\begin{figure}[t]
    \centering
    \includegraphics[width=0.45\textwidth]{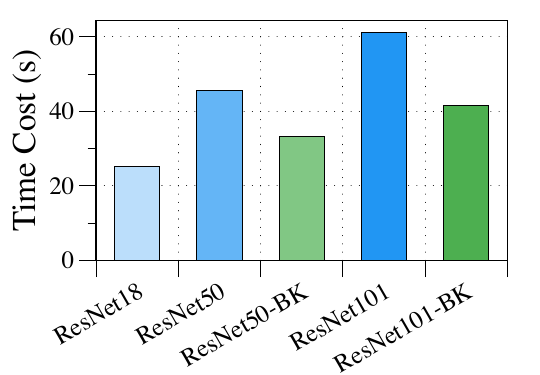}
    \caption{Ablation study on different model scales}
    \label{fig:ablation_model_scale}
\end{figure}

\subsubsection{Time cost for different model scales.}
We evaluate the time cost of \Tech{} for different model scales. Results are shonw in Figure~\ref{fig:ablation_model_scale}. For larger models, e.g., ResNet-101, \Tech{} reduces ASR to below 2\% with only a 3\% accuracy drop. The time cost increases with model size (blue bars), but even for ResNet-101, \Tech{} completes in about 1 minute.
To enhance efficiency, we can optimize only on \textit{key layers}, e.g., each residual block in ResNets. This adaptation reduces the time cost by 25\% (green bars) while still being effective.
This demonstrates \Tech{}'s efficiency even for large-scale models.

\begin{table*}[t]
    \centering
    \caption{Ablation study on different design choices}
    \label{tab:ablation_design}
    \tiny
    \tabcolsep=3.7pt
    \begin{tabular}{lgrgrgrgrgr}
        \toprule
        \multirow{2}{*}{\textbf{Attacks}} &\multicolumn{2}{c}{\textbf{No Defense}} &\multicolumn{2}{c}{\textbf{Current Setting}} &\multicolumn{2}{c}{\textbf{Act. Rejection}} &\multicolumn{2}{c}{\textbf{First Two Layers}} &\multicolumn{2}{c}{\textbf{Last Two Layers}} \\\cmidrule{2-11}
        & \cellcolor{white}{Acc.} & ASR & \cellcolor{white}{Acc.} & ASR & \cellcolor{white}{Acc.} & ASR & \cellcolor{white}{Acc.} & ASR & \cellcolor{white}{Acc.} & ASR \\
        \midrule
        BadNets &94.82\% &100.00\% &92.48\% &0.78\% &90.52\% &1.97\% &92.88\% &3.19\% &90.74\% &1.42\% \\
        Trojan &94.73\% &100.00\% &92.38\% &2.17\% &90.93\% &4.67\% &91.88\% &5.29\% &90.95\% &2.48\% \\
        Reflection &93.29\% &99.33\% &91.44\% &6.63\% &90.77\% &23.89\% &91.39\% &74.80\% &90.43\% &7.95\% \\
        Instagram &94.62\% &99.59\% &91.43\% &4.98\% &91.10\% &12.03\% &91.93\% &78.40\% &91.34\% &7.17\% \\
        DFST &93.25\% &99.77\% &91.64\% &4.02\% &91.39\% &8.62\% &92.14\% &76.58\% &90.99\% &34.76\% \\
        Adap-Blend &94.22\% &82.80\% &90.84\% &15.03\% &90.82\% &39.93\% &91.84\% &78.69\% &90.39\% &57.76\% \\
        \midrule
        Average &94.26\% &98.39\% &91.77\% &3.20\% &90.92\% &15.19\% &92.01\% &52.83\% &90.81\% &18.59\% \\
        \bottomrule
    \end{tabular}
\end{table*}

\subsubsection{Comparison Between Clipping and Rejection.}
We conduct a comparison of two approaches within \Tech{} for handling maliciously large activation values after the distribution approximation, i.e., clipping and rejection. Clipping reduces large values to the distribution boundary value, while rejection directly sets outlier values to zero. Our experiment is performed on CIFAR-10 and ResNet-18, with results presented in the first half of Table~\ref{tab:ablation_design}.
The first column denotes different attacks, Columns 2-3 present the attack performance without defense, Columns 4-5 denote the defense performance of the current setting of \Tech{} (activation clipping), and Columns 6-7 show the performance using activation rejection. Notably, clipping generally provides superior performance, resulting in higher accuracy and lower ASR compared to rejection. The underlying reason is that rejection, similar to neuron pruning, is coarse-grained. Specifically, for neurons responsible for extracting both benign and backdoor features, rejection harms accuracy while not rejecting retains the backdoor effect. In contrast, clipping eliminates only the higher values while allowing the extraction of benign features.

\subsubsection{Comparison of Operating on Different Layers.}
We examine the impact of applying \Tech{} to different layers: (1) All layers (current setting), (2) Only the first two layers, and (3) Only the last two layers. The results are presented in the last half columns of Table~\ref{tab:ablation_design}.
Observations indicate that applying \Tech{} to all layers, the current setting, generally yields the best performance compared to operating only on the first two or last two layers.
Notably, for simple backdoors such as BadNets and Trojan, where most features are extracted in the first few layers, applying \Tech{} to the first two layers is sufficient to eliminate the backdoor effect while preserving clean accuracy. However, for complex backdoors like Reflection and Instagram, where backdoor features are extracted in later layers of the network, applying \Tech{} to the last few layers achieves better performance.
Additionally, advanced attacks such as DFST and Adap-Blend, which tend to hide backdoor extraction across almost all layers, can only be effectively defended against by applying \Tech{} to all layers.

\begin{table}[t]
    \centering
    \tiny
    \caption{Ablation study of different customized accuracy degradation.}
    \label{tab:ablation_degrade}
    \begin{tabular}{lcc}
         \toprule
         \textbf{Degradation} & \textbf{Accuracy} & \textbf{ASR} \\
         \midrule
         No Defense & 94.73\% & 100.00\% \\
         \midrule
         0.01\% & 94.08\% (-0.65\%) & 4.01\% \\
         0.1\%  & 93.89\% (-0.84\%) & 3.56\% \\
         2\%    & 92.38\% (-2.35\%) & 2.17\% \\
         5\%    & 88.36\% (-6.37\%) & 1.40\% \\
         10\%   & 84.21\% (-10.52\%) & 1.12\% \\
         \bottomrule
    \end{tabular}
\end{table}

\subsubsection{Effect of Setting Different Accuracy Degradation.} \label{sec:appendix_abl_acc_deg}
We examine the impact of different accuracy degradation expectations for \Tech{}. In our experiment, we assess 5 accuracy degradation values (default is 2\%). The evaluated attack model is trained on CIFAR-10 using ResNet18 and injected with the Trojan~\cite{trojannn} backdoor.
The results in Table~\ref{tab:ablation_degrade} indicate that as the degradation increases, both accuracy and ASR decrease. This reveals a trade-off between sacrificed clean accuracy and remaining ASR.
Users seeking high clean accuracy with some tolerance for backdoor may opt for a low accuracy degradation, and vice versa.

\begin{table}[t]
    \centering
    \tiny
    \tabcolsep=7pt
    \caption{Ablation study of different optimization steps and learning rates.}
    \label{tab:ablation_step_lr}
    \begin{tabular}{lGrrrr}
         \toprule
         \textbf{Config} & No Defense & 10 \& 0.01 & 10 \& 0.001 & 50 \& 0.01 & 50 \& 0.001 \\
         \midrule
         BA & 93.51\% & 91.79\% & 91.31\% & 92.14\% & 91.41\% \\
         ASR & 100.00\% & 4.76\% & 5.49\% & 3.82\% & 4.91\% \\
         \bottomrule
    \end{tabular}
\end{table}

\subsubsection{Effect of Different Optimization Steps and Learning Rates.}
We study the effect of different optimization steps ($S$) and learning rates ($\eta$). The evaluated attack model is trained on CIFAR-10 using ResNet18 and injected with the Blend~\cite{blend} backdoor. The results are shown in Table~\ref{tab:ablation_step_lr}, where ``10 \& 0.01'' means $S=10$ and $\eta=0.01$.
We observe that \Tech{} demonstrates consistently good performance across various reasonable parameter settings, showcasing its robustness and efficiency in parameter tuning.

\end{document}